\documentclass[manuscript]{aastex}

\newcommand{\muas}{\,\mu {\rm as}}

\shorttitle{ }
\shortauthors{Sandage \& Saha}

\begin{document}


\title{Bias Properties of Extragalactic Distance Indicators.XI. Methods 
to Correct for Observational Selection Bias for RR Lyrae Absolute  
Magnitudes from Trigonometric Parallaxes Expected from the {\it FAME} 
Astrometric Satellite } 


\author{Allan Sandage}
\affil{The Observatories of the Carnegie Institution of Washington
        813 Santa Barbara Street, Pasadena, CA, 91101}

\and

\author{A. Saha}
\affil{National Optical Astronomy Observatories, 950 North Cherry Avenue, 
            Tucson, AZ 85726-6732; saha@noao.edu}




\begin{abstract}
     A short history is given of the development of the 
correction for observation selection bias inherent in the 
calibration of absolute magnitudes using trigonometric 
parallaxes. The developments have been due to Eddington, 
Jeffreys, Trumpler and Weaver, Wallerstein, Ljunggren and Oja, 
West, Lutz and Kelker after whom the bias is named, Turon 
Lacarrieu and Cr\'{e}z\'{e}, Hanson, Smith, and many others.
     As a tutorial to gain an intuitive understanding of several complicated 
trigonometric bias problems, we study a toy bias model of a parallax catalog 
which incorporates assumed parallax measuring errors of various severities. 
The two effects of bias errors on the derived absolute magnitudes are (1) the 
Lutz-Kelker correction itself that depends on the relative parallax error 
$\delta \pi / \pi$ and the spatial distribution, and (2) a Malmquist-like 
`incompleteness' correction of opposite sign due to various apparent 
magnitude cut-offs as they are progressively imposed on the catalog.
     We calculate the bias properties using simulations 
involving $3 \times 10^{6}$ stars of fixed absolute magnitude using 
$M_{v} = +0.6$  to imitate RR Lyrae variables in the mean. These stars are 
spread over a spherical 
volume bounded by a radius 50,000 parsecs with different spatial density 
distributions. The bias is demonstrated by first using a fixed 
rms parallax uncertainty per star of $ 50 \muas$ , and 
then using a variable rms accuracy that ranges from $ 50 \muas $ 
at apparent magnitude $V = 9$ to $500 \muas$ at 
$V = 15$ according to the specifications for the {\it FAME} astrometric 
satellite to be launched in 2004. The effects of imposing magnitude limits 
and limits on the `observer's' error, $\delta \pi / \pi$, are displayed.
     We contrast the method of calculating mean absolute magnitude directly 
from the parallaxes where bias corrections are mandatory, with an inverse
method using maximum likelihood which is free of the Lutz-Kelker bias, 
although a Malmquist bias is present. Simulations show the power of the 
inverse method. Nevertheless, we recommend reduction of the data using both 
methods. Each must give the same answer if each is freed from systematic error.
Although the maximum likelihood method will, in theory, eliminate many of 
the bias problems of the direct method, nevertheless the bias corrections required by the direct method can be determined {\it empirically} via 
Spaenhauer diagrams immediately from the data, as discussed in the earlier 
papers of this series.  Any correlation of the absolute (trigonometric)
magnitudes with the (trigonometric) distances {\it is the bias}.  
      We discuss the level of accuracy that can be expected in a calibration of RR Lyrae absolute magnitudes from the {\it FAME} data over the metallicity range of $ {\rm [Fe/H]} $ from $0$ to $-2$, given the known frequency of the local RR Lyraes closer than 1.5 kpc.  Of course, use will also be made of the 
entire {\it FAME} database for the RR Lyrae stars over the complete range of distances that can be used to empirically determine the random and systematic errors from the {\it FAME} parallax catalog, using correlations of derived 
absolute magnitude with distance and position in the sky. These bias corrections are expected to be much more complicated than only a function of apparent 
magnitude because of various restrictions due to orbital constraints on the 
space-craft. 
     
\end{abstract}




\section{INTRODUCTION}

     The approved NASA mission, {\it FAME}, is a science program based 
on an astrometric satellite that will obtain all-sky 
trigonometric parallaxes and proper motions for stars brighter 
than magnitude 15. The accuracy of the parallaxes has been 
specified to be $ 50 \muas$  in the best range of the 
space-craft's configuration for stars brighter than $V = 10$ mag, 
degrading to no worse than $500 \muas$ at its detection 
limit at $V = 15$. These accuracies are between 2 and 20 times more 
accurate than achieved by Hipparcos. The typical rms accuracy for 
trigonometric parallaxes of Hipparcos data is $1000 \muas$, 
spectacular at the time but not accurate enough by at 
least a factor of 10 to reach the domain of the RR Lyrae stars. 
The promise of the data from {\it FAME}, if its mission goals are 
achieved, is that the domain needed for the RR Lyrae absolute 
magnitude calibration can be achieved directly via trigonometric 
parallaxes. 

     The purpose of the present paper is to assess what must be 
done with the RR Lyrae parallax data from {\it FAME} in order to 
correct for observational selection bias in determining a correct 
calibration of absolute magnitude for these stars as a function 
of their metallicity. 

     It is widely recognized that (1) the solution to many 
problems in Galactic structure, (2) an account of the episodes 
and time-sequences in the formation of the Galaxy, and (3) one 
approach to the extra-galactic distance scale, rest directly on 
the calibration of $M_{V}(RR) = f({\rm [Fe/H]})$. In particular, the 
steepness of the dependence of $M_{V}(RR)$ on ${\rm [Fe/H]}$ determines 
whether there is an appreciable time interval over which the 
Galactic globular clusters of different metallicities have 
formed, or whether the entire Galactic globular cluster system 
formed nearly simultaneously with the rapid collapse of the 
nascent Galaxy with its early separation of the disk and the halo 
(Baade 1957; Eggen et al. 1962; Sandage 1986, 1990a). 

     It has been shown elsewhere (Sandage and Cacciari 1990) that 
if the slope of the relation between absolute magnitude and 
metallicity for RR Lyrae stars is as large as 
$dM_{V}(RR)/d([Fe/H]) = 0.32$, 
then there is no dependence of the ages of the Galactic 
globular clusters on $[Fe/H]$. This conclusion was based on the 
stellar models available in 1990, before the $[O/Fe]$ enhancement 
was known. However, the same dependence of the globular cluster 
formation history on the value of the $dM_{V}(RR)/d([Fe/H])$ slope has 
more recently been confirmed by Chaboyer, Demarque, and 
Sarajedini (1996) using Oxygen enhanced models. 

     The dependence of the age spread on the slope was 
not discussed by Chaboyer {\it et al.}  They only treated the 
$dM_{V}(RR)/d([Fe/H]) = 0.20$ case, but their large age spread 
for this low-slope case is nearly identical with that of Sandage 
and Cacciari (1990) for the same small slope. 
If they had used the steeper metallicity dependence that is 
required by the observed Oosterhoff period effect (Sandage 
1993a,b), their conclusion concerning a large age spread 
depending on $[Fe/H]$ would have been reversed. No age spread is 
predicted if the preferred slope of 0.30 is used. In fact, the 
Oxygen enhanced models of Bergbush and Vandenberg (1992) show 
that if the slope is as shallow as $dM_{V}((RR))/([Fe/H]) = 0.26$, 
then there is no age spread among the globular clusters of 
different metallicities. Clearly, one crucial importance of the 
{\it FAME} mission will be its ability to determine a definitive 
calibration of $ M_{V}(RR) = f([Fe/H]) $ relation for RR Lyrae stars. 
The data impact directly on the formation history of the Galaxy. 

     However, impressive as  the specified accuracy of $50 \muas$ is 
for the parallax accuracy for stars at $V = 10$ mag, 
that accuracy is near the margin of what is needed to make a 
definitive calibration of $M_{V}(RR)$ as a function of metallicity. 
The uncertainties center on the inevitable observational 
selection bias due to the distribution of parallax errors that 
will exist in the highly non-linear distribution of the observed 
parallaxes. This effect is currently named the Lutz-Kelker bias. 
The problem has a long 
history, part of which we review in the next section.

\section{A SHORT HISTORY OF A BIAS IN ABSOLUTE MAGNITUDES DUE TO 
         MEASURING ERRORS IN TRIGONOMETRIC PARALLAXES}

\subsection{The beginnings}

     Discussions of the correction needed to recover a true 
distribution from an observed distribution in the presence of 
observational errors began at least as far back as 1913.         
Eddington (1913), as chief assistant to Sir Frank Dyson, 
then Astronomer Royal of England, derived an equation by which to 
recover the true distribution of some particular measured 
quantity, such as the number of stars with particular measured 
parallaxes, from the observed values which have a distribution of 
measuring errors. The method not only recovered an approximation 
to the true distribution of the measured parallaxes but also 
gave, in a later expanded form, an estimate of the difference 
between the observed and the true mean value of the parallax for 
particular binnings of the data.
     
     Eddington's theoretical solution lay dormant until Dyson 
(1926) took it up in his discussion on how to treat negative 
parallaxes caused by observational errors. The problem reduced to 
how to derive the true distribution of parallaxes from the 
observed distribution, estimating the mean rms error from the 
negative tail of the observed parallax distribution (e.g. Smart 
1936, sections 1.81, 1.82; Nassau 1928a,b). 

     Dyson's discussion using Eddington's method was criticized 
by Jeffreys (1938) in a remarkable paper where at one point he 
states; ``These restrictions [on the conditions of the 
distributions] are so severe that I doubt whether the method 
could ever be correctly applied in practice, and in any case 
better methods exist.'' The better method alluded to here by 
Jeffreys was even at that time called ``the method of maximum 
likelihood" of R.A. Fisher (1912, 1925), following embryos of the 
method as they had been developed by Gauss, Helmert, and Pearson. 

     The maximum likelihood program treats the problem by 
interchanging the independent and dependent variables of the 
system. We shall encounter the same alternate ``inverse" method in 
\S 6 of this paper, following Gould's suggestion to us on 
the inverse method of solution based on the maximum likelihood 
program. It is, of course, interesting that even in 
1938, the direct and inverse methods were beginning to be argued 
by the giants that walked the Earth in those days.      

     The most important paper after those of Eddington, Dyson, 
and Jeffreys was again by Eddington (1940). There he elaborates 
on his 1913 paper, derives the correction to the mean value of an 
observed distribution due to bias, and answers the attack by 
Jeffreys. Much of the modern literature on the bias rests on the 
foundations of this paper. 

     For the purposes of understanding the systematic bias in the 
mean value of the observed distribution function compared with 
the true mean, the exceptionally clear statement of one cause of 
the bias is made by Jeffreys (1938): 

    ``A series of quantities are measured and classified in equal 
ranges. A measure has a known uncertainty. On account of the 
errors of measurement some quantities are put into the wrong 
ranges. If the true number in a range is greater than those in 
the adjacent ranges, one should expect more observations to be 
scattered out of the range than into it, so that the observed 
number will need a positive correction" 

     This cause of the bias in an absolute magnitude calibration 
is mentioned by Trumpler and Weaver (1953), using an example of a  
sample of stars with observed parallaxes of 0.02 arc-sec. They 
write:  

     ``Errors of observation will vitiate [the statistical 
properties of the sample] in two ways. \\
~~   (1) The number of stars with observed parallax values greater 
than 0.02 arc-sec is not the true number of stars with distance smaller 
than 50 parsecs. Many stars having a true parallax smaller than 
0.02 arc-sec will be erroneously included among the stars in the sample 
volume because the measured value [of the parallax] is too large. 
Some stars with a true parallax larger than 0.02 arc-sec will be 
omitted because the result of observation is too small. In 
general, however, the omissions will not cancel the additions; 
the latter will usually be more numerous [because there are more 
stars with true distances larger than 50 parsecs than with 
smaller distances unless there is a steep density gradient 
outward steeper than $\rho = d^{-2}$ ]. \\
~~    (2) When stars are selected by a lower limit in the observed 
parallax value, [i.e. if the sample is arbitrary cut off at some 
distance limit], we favor stars for which the measured parallax 
result is too large. The absolute magnitudes calculated with the 
observed parallax values will thus be systematically too large 
[i.e. {\it too faint}]."  

     Item (2) introduces the notion that the effect on the 
calibrated mean absolute magnitude, $\langle M \rangle$, depends on the placing 
of a lower limit to the observed parallax distribution. The 
advance made by Lutz \& Kelker (1973) was the showing that this 
restriction of placing a lower parallax limit is not responsible 
for the existence of the bias. Many of the early discussions of 
the bias were based on placing a lower parallax limit to the 
observations. Lutz \& Kelker showed that decision on a lower 
limit to be irrelevant to the existence of the bias.    

      The Jeffreys/Eddington exchange, the Trumpler/Weaver 
paragraphs, and the application of the Eddington theory to the 
actual case of the absolute magnitude by Nassau (1928a,b) in 
general, and by Feast and Shuttleworth (1965) for B stars, 
brought to a close the beginning period of the problem in its 
development in the literature. 

\subsection{The middle period}

     The next advance came in a unexpected way via what at 
appeared at first to be an unrelated route. In a remarkable, 
highly prescient, and eventually hotly debated paper, Hodge and 
Wallerstein (1966) proposed that the distance modulus of the 
Hyades should be increased from its canonical value of 3.03 mag 
to $ m-M = 3.42 $. Their arguments were based on the properties of 
the stellar models for mass and luminosity of main sequence 
stars. 

     The consequences of such an increase in the Hyades distance 
would be felt across the entire subject of stellar astronomy from 
stellar evolution to observational cosmology. The Hyades main 
sequence had been taken to define the age-zero main sequence to 
which all photometric parallaxes were tied at the time. One of 
the arguments used by Hodge and Wallerstein centered on O.C. 
Wilson's (1967) absolute magnitude calibration of the Ca II H and 
K line-width-absolute magnitude correlation which he had 
discovered and had advanced with Bappu (Wilson and Bappu 1957).   

     By various arguments, Hodge and Wallerstein made a case that 
Wilson's calibration for giants must be systematically in error 
(too faint) by perhaps as much as 0.5 mag. This calibration was 
based on trigonometric parallaxes for giants whose observational 
errors were not small compared with the measured parallax. 

     This suggestion by Hodge and Wallerstein seemed outrageous 
to many critics. Heavy criticisms of the Hodge-Wallerstein paper, 
almost all of which by hindsight are now seen to be unjustified, began to 
appear. However, rather than fight the critics, who he knew to 
be largely wrong, Hodge left the problem so as to enjoy himself 
in his other productive Elysian fields which the incorrect 
critics failed to find. 

     On the other hand, Wallerstein, also confident that a 
systematic error must exist in the trigonometric parallax values 
for the four calibrating giants used by Wilson, discovered the 
Eddington/Trumpler/Weaver bias in the literature and set out to 
understand it in practical (operational) terms. His paper 
(Wallerstein 1967) began the modern era for this trigonometric 
bias problem.    

     Wallerstein started with the statements of Jeffreys, and of 
Trumpler and Weaver about the asymmetry between stars leaving the 
various parallax ranges and those entering the ranges.  
Wallerstein quantified the effect of the asymmetry by using the 
new powerful statistical method of Monte Carlo simulations. The 
bias showed immediately from his simulations at the level of 0.5 
to 0.8 mag. This was just the level that Hodge and Wallerstein 
had predicted from their astrophysical arguments.       

     Following Trumpler and Weaver's point (2), Wallerstein had 
put a lower limit at 0.007 arc-sec to his artificial parallax 
catalogue of 4096 stars. He then followed the effect of the 
asymmetry in the ratio of incoming stars to outgoing stars in 
bins of measured parallax because of measuring errors. 

     Wallerstein gives only a summary of the results, but clearly 
states the direction of the bias. The {\it true} absolute magnitude 
calibration of a sample of parallax stars is brighter than the 
direct calculation of the mean absolute magnitude of a sample, 
calculated using the {\it measured} parallaxes. Wallerstein not only 
applied his results to the Wilson calibration of the Wilson-Bappu 
effect, but also to the calibration of the position of the age-
zero main sequence which Eggen and Sandage (1962) had derived 
from trigonometric parallaxes as small as 0.035 arc-sec in the 
presence of individual parallax errors that were as large as 
0.0065 arc-sec. 

     Wallerstein obtained corrections to the position of the main 
sequence that ranged between 0.12 and 1.03 mag, depending on the 
parallax. Clearly, this Eddington effect, so called at that time, 
was a major problem that had not been dealt with in these 
contexts before. Just how serious was it? In the case of the main 
sequence position of subdwarfs, the correction impacted directly 
on the age of the globular clusters (Sandage 1970).   

     In the context of the calibration of the Wilson Ca II H/K 
effect for giants, the next important paper on the bias is that 
of West (1969). This eventually became more widely known than a 
similar, earlier paper by Ljunggren and Oja (1965), that had, in 
fact, preceded Wallerstein's (1967) simulations. These two 
papers, one by Ljunggren and Oja and the other by West, used 
analytical methods, following Eddington, to derive the bias 
properties of parallax samples using standard methods of  
statistical astronomy to calculate mean values of observed 
distributions from the known rms measuring errors.   

     The West paper was the clearest description of the method at 
the time. West also gave an important table of results. These 
have been repeated and verified in all subsequent papers on the 
problem. His summary table contains the calculations of the bias 
offsets in magnitudes for a range of values of the fractional (or relative)
parallax error, $ \delta \pi / \pi $, for various distributions of the 
true parallaxes. The steeper is the true parallax distribution, 
(i.e. the number of parallax entries as a function of the 
measured parallax in unit parallax interval) as the spatial 
density of the sample increases outward, the more candidate stars 
will be erroneously thrown into the observed ranges from larger 
distances than thrown out from smaller distances. Clearly, the 
bias error due to the asymmetry will be larger for the steeper 
spatial density distributions. 

     The next advance was made by Lutz and Kelker (1973) who used 
the formalism of West but treated only the constant density 
case. There, the distribution of the true parallaxes will be 
$ N(\pi) d \pi ~=~ (4 \Pi \rho d \pi)/(\pi^{4}) $\footnote{in this paper we
use the symbol $\pi$ or $p$ to denote parallax, and $\Pi$ to denote the ratio 
of circumference to diameter of a circle.}
 (see \S 3.2), where $N(\pi)$ 
is the number of 
stars with parallaxes between $ \pi $ and $ \pi + d \pi $, and $\rho$ 
is the spatial density of stars. Whereas West 
treats a series of cases with variable spatial density, Lutz and 
Kelker only treat the constant density case where the exponent on 
$\pi$ in the parallax distribution function is $n = 4$ as above. 
West treats the cases also with $n = 3$ and $n = 2$.
 
     The first advance made by Lutz and Kelker over West (and 
indeed Trumpler and Weaver in their item 2) is that they show 
that the presence of the bias does not depend on making a lower 
limit to the sample, but rather is present at all levels of the 
parallax distribution. Their second advance was to cast the 
problem into dimensionless form, emphasizing that the size of the 
effect depends only on the relative parallax error, ($ \delta \pi / \pi $).
 
     The Lutz-Kelker paper was so clear that it soon became the 
principal reference to the problem, even as Wallerstein, West, 
and Ljunggren and Oja (1965) had derived the same results 
earlier. The methods of both Ljunggren and Oja, and 
West were analytical, based on the classical equations of statistical 
astronomy. The simulation method of Wallerstein was new and 
eventually proved to be very powerful in elucidating the problem 
in a highly intuitive way. We follow Wallerstein's method in our 
simulations in \S 3. 

\subsection{The present period}

     Following Lutz and Kelker the literature began an 
exponential expansion. We mention here only a few of the many 
papers, both analytical and practical, many of which provide 
illuminating summaries of the problem.
 
     In a prescient paper, Turon Lacarrieu \& Cr\'{e}z\'{e} (1977)           
repeat the analysis of West, and of Lutz and Kelker, showing 
excellent agreement (their Table 3) in the magnitude of the bias 
correction for various values of $ \delta \pi / \pi $. They also give the 
alternate solution of the problem by exchanging the independent 
and dependent variables (the parallax error and the intrinsic 
variation of the absolute magnitude itself). This is the inverse 
method of ``maximum likelihood", which, if applied properly, has no 
selection bias to the derived mean absolute magnitude (see section 
6 here). This is one of the ``better methods" mentioned by 
Jeffreys, and due to R.A. Fisher (1912, 1925), as set out in 
\S 2.1. Turon, Lacarrieu \& Cr\'{e}z\'{e} apply the method to 
recalculate the calibration of various intermediate-band 
Str\"{o}mgren indices and to compare their maximum likelihood 
calculation with the calibration of Crawford (1975) who used the 
Lutz-Kelker correction directly. 

     Lutz (1978) discusses the bias by a clear elementary 
example. He also summarizes the main result of the Lutz-Kelker 
(1973) analytical calculation. In an important paper, Lutz (1979) 
reanalyzes the problem if a restriction is placed on the apparent 
magnitude of the sample. He also gives a later summary (Lutz 1983). 
The magnitude restriction is a crucial part of the problem, as will 
become evident in \S 3 from our own simulations. This becomes the 
governing aspect of the bias correction at large distances, actually 
changing the sign of the magnitude correction.

     A series of papers on the application of the bias correction 
to problems other than the original correction to O.C. Wilson's 
calibrations of his H/K indices began to appear in 1979. 
Perhaps the most important of these is that of Hanson 
(1979). There he removes some of the uncertainties of the method 
that requires knowledge of the true parallax distribution (the 
beta exponent of West, renamed here as $n+1$ in section 2.2 above). 
He does this by appealing to the proper motion distribution 
that does not contain the same type of bias. 

     Hanson's main observational discussion concerns the required 
Lutz-Kelker correction to the trigonometric subdwarfs used by 
Sandage (1970) to which to fit the globular cluster main 
sequences. Here, he analyzes in greater detail than was done by 
Wallerstein (1967) for the main sequence position derived by 
Eggen and Sandage (1962), mentioned earlier (\S 2.2). The summary 
paper given by Koen (1992) is also important to cite. 

     The most recent detailed discussion of the correction to 
subdwarf parallaxes is by Reid (1997) where the Lutz-Kelker 
correction is summarized and is applied to each of the new 
subdwarf parallax values from Hipparcos. There, the West/Hanson 
values of the correction as a function of the space density 
parameter, $n$, is compared with an analytical formula due to Smith 
(1987c) that applies to the case of $n = 4$. The paper by Smith is 
the last of a series by him (Smith 1987a,b) that illuminates the 
problem, including the inverse case using the maximum likelihood 
method.  

          The literature on the bias problem again began a major  
expansion as soon as the Hipparcos parallax catalog appeared. 
Some of the initial analyses of absolute magnitude calibrations 
used the direct method, often with only a passing mention of 
the bias problem, and also often using no corrections at all. 
Many other papers did made efforts to correct for the bias but 
solely on the basis of the analytical models such as those by 
West, Hanson, Smith, Koen, and of course Lutz-Kelker themselves. 
However, in each case this required a decision for the appropriate 
spatial density distribution in order to enter the analytical 
tables. 
     Warnings on the blind application of bias corrections to the 
Hipparcos data were well set out by Brown et al. (1997) in their 
paper on properties of the Hipparcos catalog. Their explicit 
recommendation for the Hipparcos database will apply also to 
analyses of the {\it FAME} data. They write:

   ``The reader is strongly encouraged to perform a detailed 
analysis [of the sort outlined here] {\it for each specific case} in 
order to obtain a correct estimation of any parameter of a star 
or a sample of stars using trigonometric parallaxes. This means 
in particular that one should neither ignore the possible biases 
nor apply blindly `Malmquist' or `Lutz-Kelker' corrections.''

      The important paper by Reid (1997) on the calibration of the 
subdwarf main sequence as a function of metallicity is 
a case in point. He uses the analytical models of Hanson (1979) 
and of Smith (1987c), with a choice of the spatial density 
parameter. His adopted bias corrections based on Hanson's fitting 
equation, ranged from $0.00$ mag to $0.42$ mag, corresponding to 
relative parallax errors of $ \delta \pi / \pi$ between $0.007$ and $0.196$. 
The largest corrections carry an uncertainty of a factor of two 
depending on which of the possible density distributions is 
chosen.

     It is this uncertainty of the analytical models that is the 
final message of the present paper. Although the analytical and 
simulated models often lead to a more adequate understanding of 
the bias problems, they do not provide the necessary accuracy 
because of their strong dependence on the variety of input 
parameters (spatial density distribution and apparent magnitude 
cut offs). For this reason, we shall later advocate in sections 
3.2.7, 5, and 6 that if the direct method of calibration is used, 
the bias corrections should be determined {\it empirically} from 
the embedded data in the database itself using such methods as 
Spaenhauer diagrams, for example, explored in earlier papers of 
this series.

\section{THE SIMULATIONS OF THE EXPECTED ACCURACIES OF 
$M_{V}(RR) = f(\rm{[Fe/H]})$ FROM THE ``{\it FAME}'' MISSION}

\subsection{The purpose of this paper}

    Our purpose is to pursue the method of simulations started 
by Wallerstein (1967) in order to develop a better intuitive 
understanding of why an Eddington/Jeffreys/West/Lutz-
Kelker/Turon et al./Hanson/Smith/etc. bias occurs in 
trigonometric parallaxes. To that purpose, we have begun anew, at 
the pioneering level of Wallerstein (1967), to simulate the 
selection bias due to the finite errors in the measured 
parallaxes. We wish to understand the bias from the practical 
approach of an observer confronted with a catalog of parallaxes 
that has the individual rms uncertainties listed for each catalog 
star.  
 
     Hence, this paper is not, {\it per se}, a methods paper, setting 
out any detailed procedure on how astronomers will eventually use 
the vast database that is expected from the {\it FAME} space-craft. 
Rather, our purpose is simply to proceed step by step, 
complication added to complication, to understand the reason for 
bias in the calibration of mean absolute magnitudes for any 
particular class of stars when using a parallax catalog such as 
will be produced by {\it FAME}. 

     To initially gain a better intuitive understanding of the 
reasons for any bias, we first construct a toy model of a 
parallax catalog where every star has the same error in its 
measured parallax of $ 50 \muas $, independent of its apparent 
magnitude. It is, of course, to be understood that this first toy 
model is unrealistic in many ways. For example, there will be a 
spread of errors in addition to those that depend on apparent 
magnitude. One addition to the error will be a dependence on 
ecliptic latitude due to the severe restrictions on the duty 
cycle (the number of times revisited) caused by restrictions on 
positions relative to the sun. Hence, the parallax error will be 
differ, star-by-star, and position-by-position even at a given 
apparent magnitude.  
 
     Therefore, we contend that the error for any given star is a 
complicated enough function of many parameters, in addition to the 
apparent magnitude, that the empirical approach to the bias 
corrections advocated later (\S 6) is safer than any 
analytical approach via models with idealized input parameters. 
Our purpose in using such idealized toy models, is simply to 
understand the effect of these input parameters so that we can 
gain an intuition of what the bias problem is about in the real 
cases that the {\it FAME} databases will present. If we cannot 
understand these simplified cases, we probably will not 
sufficiently understand reality when it is presented by the {\it FAME} 
data.

\subsection{The problem to be solved}
    
     After setting out the toy model in \S \S 3.3.1 to 3.3.5 and its more 
realistic form in \S 3.3.6, we proceed to assess the accuracy 
with which the absolute magnitude calibration of RR Lyrae 
stars as a function of metallicity can be obtained from the {\it FAME} 
trigonometric parallaxes if the proposed specifications of the 
satellite are met. Those specifications are that the rms errors 
of parallaxes for individual stars are to be $24 \muas$ at magnitude 
$V = 9$, ~$36 \muas$ 
at $V = 10$, ~$56 \muas$  at $V = 11$, ~$90 \muas$
at $V = 12$, rising to $250 \muas$ at $V = 14$. 

     There are two parts to the analysis. (1) Using an 
assumed mean absolute magnitude of $ +0.6$ mag for RR Lyrae stars, 
consistent within the range of most modern calibrations, we first 
calculate the parallax bias corrections for RR Lyrae stars 
between V = 9 and 12. Here we use both a constant rms parallax error 
(\S3.3.1 to \S3.3.5),
and then a variable rms error 
(\S 3.3.6) with magnitude. The ``observer's" distances of 
such stars between these apparent magnitude limits are between 
480 and 1620 parsecs. (2) We determine in \S 6 if there 
are enough RR Lyrae stars in this distance range to reduce the 
rms spread about the mean bias correction in order to produce a calibration 
of the mean absolute magnitudes in say three bins of metallicity 
between $ {\rm [Fe/H]} = 0$ and $-2.5$.
 
     We approach problem (1) by the method of simulations 
pioneered by Wallerstein. The object is to find the bias 
directly, first in \S 3.3.1 for the unrealistic but simple 
case of constant rms error of $50 \muas$ at all apparent 
magnitudes, and then in \S 3.2.2 in the more realistic case of varying 
the rms accuracy according to apparent magnitude taken from the 
specifications set for the {\it FAME} satellite.
 
     We address problem (2) in \S 6 by determining the actual 
number of RR Lyrae stars that the {\it FAME} catalog will contain in 
each distance interval and in each metallicity bin by counting 
the RR Lyrae population of given metallicity in a standard 
variable star catalog. The uncertainty in determining the bias correction 
for each metallicity bin is estimated by attenuating the rms spread about 
the mean bias correction by the square root of the number of RR Lyrae stars 
in the bin. 

\subsection{The simulations}

\subsubsection{The simplistic case of a constant rms accuracy at all distances}

     The first group of simulations is made by distributing $3 \times 10^{6}$
stars in a volume bounded by a maximum distance of 50,000 
parsecs. We consider three spatial density distributions where 
the {\it cumulative} number of stars, $N(R)$, enclosed within a distance 
$R$ varies as $R^n$, with $n = 3, 2,$ and $1$. These integral 
distributions in {\it distance} are identical with the $\beta$ exponents 
of 4, 3, and 2 in West's (1969) formulation using the 
differential distribution of {\it parallaxes}, $f(p) dp$ where $f(p)$ is the 
number of stars with parallax $p$ in parallax interval $dp$ (see 
below).   

     Each star is assigned a true distance based on the assumed 
fixed absolute magnitude of $<M_{V}> = +0.6$, characteristic 
of RR Lyrae stars near $[Fe/H] = -1.2$. Hence, each star has the 
true apparent magnitude of $m = 5 \log D + 4.4 $. 

     Each is then given a random parallax error drawn from a 
Gaussian distribution of the errors with an rms value of $50 \muas$. 
This simulated catalog changes the true 
catalog into the ``observer's" catalog by the introduction of the 
rms random parallax error. The ``observer's" catalog is then used 
to calculate the mean absolute magnitude that an observer would 
obtain using the measured parallaxes that carry the rms errors. 
Each derived absolute magnitude will differ from $M = +0.6$ by the 
amount that the observed parallax differs from the true parallax 
by the measuring errors.   

     The systematic magnitude bias is the difference between $M = 
+0.6$ and the derived mean absolute magnitude of subsets of the 
data selected in various ways from the ``observer's" catalog. The 
bias will vary with (1) the inferred (i.e. the ``observer"s) 
distances, (2) with the $\delta \pi / \pi$ fractional parallax error, 
and (3) with any magnitude (or observed parallax) cutoff that the 
observer makes in the sample selection from the catalog. 

     The rms variation of the mean bias is also a crucial 
observed quantity. As stated earlier, this tells how many sample stars at a 
given distance, and apparent magnitude (or parallax) cut-off, are 
required to determine the mean $<\Delta M>$ bias error to within a 
given statistical error. Are there enough RR Lyrae stars in the 
sky at the various distance and metallicity ranges to reduce the 
error of $\langle M \rangle$ to an acceptable level for a useful calibration of 
$M_{V} = f([Fe/H]$? We address this question in \S 6 by 
calculating the $<\Delta M>$ bias values as a function of distance 
for various rms errors as a function of magnitude and for various 
assumed spatial density distributions, given the number of available stars in the sample.  

\subsubsection{The uniform spatial density distribution case: 
$ N(R) \sim  R^3 $ }           

      The (true) number of stars, $F(R)$, in each shell of radius $R$ 
of thickness $dR$ is the distribution function that varies with $R$ 
as $F(R) \sim R^2$. Define the distribution of (true) parallaxes to be 
$G(p)$.  This is the number of parallaxes of size $p$ in parallax 
interval $dp$. With a spatial function $F(R)$ that increases 
as $R^2$, the $G(p)$ decreases with $p$ as $p^{-4}$.
This follows because 
     
\begin{equation}
                G(p)dp = F(R)dR
\end{equation}
          
i.e the numbers are conserved between the representations using 
either {\it distance} or {\it parallax}. Because $R = p^{-1}$, then 
$dR/dp= - R^2$, 
which, when put in equation (1) with $F(R) = R^2$, gives 
\begin{equation}
              G(p)dp = F(R)(dR) = p^{-4} dp
\end{equation}
a well known result for the constant spatial density case (e.g. 
West 1969; Lutz and Kelker 1973).  

     The upper left panel of Figure 1 is a comparison of the assumed 
true (intrinsic) spatial distribution for this case of $F(R)$ with 
the apparent distribution in the ``observer's" catalog (the 
histogram) where no restriction is placed on the magnitude limit 
for a sampling in the ``observer's" catalog. The histogram is 
plotted in bins of 50 parsecs width. 

     The three remaining panels of Figure 1 show the ``observer's" 
distributions for three different apparent magnitude cuts in the 
catalog with $m$(limit) of 15, 14, and 13 mag respectively, 
discussed later in this section. 

     The parallax distributions that correspond to the spatial distribution in 
the upper left of Figure 1 is steep at $p^{-4}$, i.e. it is highly 
non-linear. Hence, if we were to run an error filter that is 
{\it symmetrical} in {\it parallax} over such a steep non-linear parallax 
distribution, we will clearly throw more stars into larger 
parallaxes (smaller distances) than we throw out. This is the message 
of Lutz (1978) in his clear, elementary example. Hence, the 
apparent spatial distribution of the ``observer's catalogue" will 
be steeper (more stars at smaller distances) than the true 
distribution. This is precisely what Figure 1, left panel, shows. 
Because there will be an excess of stars at smaller implied 
distances in the ``observer's" error catalog, compared with the 
true distribution, the consequence is that the inferred {\it mean}
absolute magnitude for such a sample of stars will be fainter 
(smaller distances for the same apparent magnitude as in a true 
catalog) than in the true sample. {\it This is the bias}. It is the 
demonstration of point (1) made by Trumpler and Weaver that was 
discussed in \S 2.
  
     Figure 2 shows a different representation. The absolute 
magnitude calculated for each star in the observer's catalog is 
plotted vs. its inferred distance, $R(observed)$. The line for the 
true absolute magnitude of $M = +0.6$ mag is shown as the white 
stripe. The statistical difference in the mean distribution of 
points relative to the white stripe is the bias, growing as a 
function of distance. 
     An important feature of Figure 2 (upper left panel) is that 
the errors for many stars become so large at about $R = 4000$ 
parsecs that any bias correction becomes unmanageable at small 
{\it measured} parallaxes. This is because the large number of stars 
that have inferred absolute magnitudes fainter than say +3 
(compared with their true value of +0.6), are in fact at very 
large true distances. The error in $R$ at distance $R$, for a given 
parallax error, $dp$, goes as the square of $R$ for this density 
distribution. The derivation is as follows.
     From $R = \pi^{-1}$, the error, $dR$ in $R$ for a given parallax 
error, $d \pi$ in $pi$ is 
\footnote{It is the $R^2$ term in equation (3) 
that transforms the assumed 
{\it symmetrical} (Gaussian) error distribution in {\it parallax} into the 
highly {\it asymmetrical} error distribution in $R$, the consequence of 
which is described in this section. It is this asymmetry that 
leads to the bias. This is the message of Lutz (1978). }

\begin{equation}
                  dR = - \pi^{-2} d \pi = - R^{2} d \pi,        
\end{equation}

     Hence, stars at large $R$ show the largest migration into the 
observable range. By restricting the apparent magnitude, as in 
three of the panels in Figures 1 and 2, we restrict the distances 
that enter into the calculation of the bias values.  
     It is clear from Figures 1, 2, and later from Figures 4, 5, 
 and 7 that the effect of the magnitude cutoff drastically 
changes the distributions of the bias error for distances larger 
than about 2500 parsecs. Hence the runaway tail at $R > 3000$ 
parsecs can be controlled by the observer if she/he will not use 
the entire ``observer's catalog", but will restrict the sample by 
apparent magnitude as in the last three panels of Figures 2, 5 and 
7. Lutz (1979) studied this case. West did a similar analysis by 
limiting the sample by putting a restriction on the observed 
parallax.     

     We quantify the effect in the three remaining panels of 
Figure 2 by restricting the total sample to subsamples with 
limiting magnitudes of 15, 14, and 13. The effect, of course, is 
dramatic. It will be paramount in the discussion of the realistic 
case in \S 3.3.2 where the rms parallax errors are assumed 
to increase with increasing faintness as in the {\it FAME} 
specifications. 

     Histograms of the detailed distribution in Figure 2 for a 
magnitude restriction at $m = 15$, binned in distance intervals of 
500 parsecs are shown in Figure 3 for the ``observer's'' distance from 
2000 to 4000 pc. The histograms for distances from 0 to 2000 pc are of the 
same form. They, of course,
have smaller bias corrections and smaller rms dispersions (Table~1), and 
are not shown, 
in order to conserve space.  These distributions are 
important because they not only show the offset of the mean line 
(dashed vertical) from +0.6 mag, which is the mean bias, but 
they also show the {\it distribution} of the residuals about this mean 
bias error. The headers for each panel give the distance range, 
the assumed rms parallax error of $50 \muas$, the mean 
absolute magnitude, the rms deviation from this error, and the 
number of stars making up the distribution.
 
     It is the rms deviations about the mean line 
that determine how accurately 
the mean bias error can be measured using only a finite number of 
stars in any actual catalog (i.e. by the actual number of RR 
Lyrae stars that are available in the Galaxy). This is the second 
problem mentioned in \S 3.2 to be discussed in \S 6. For example, 
the bias error of 0.16 mag in the upper left panel of Fig.~3 for 
the distance range of 2000 to 2500 parsecs has an rms deviation 
of 0.26 mag. If there are 66 RR Lyrae stars in this distance 
range, as in Layden's (1994) list over all metallicities, (cf. 
\S 6), the systematic bias correction of 0.26 mag can be 
determined to within an rms accuracy of 0.032 mag. However, if we 
also wish to break the sample still further into say five 
metallicity groups, the error per group will be larger at 0.07 
mag if there would be 13 such stars in each bin. The actual cases 
are more complicated when we use the projected realistic errors 
for {\it FAME} and the actual metallicity distributions in Layden's 
list set out in \S 6a. 
    
     Table~1, in columns 3 and 6, is a summary of the results 
from Figures~1-3 for the spatial density case of $n = 3$. We have 
binned the data in Figure~2 into the discrete distance intervals 
listed in column 1. The relative parallax error, $ \delta \pi / \pi$, in 
column 2 is for the midpoint of the distance interval, based on 
the fixed parallax error of $ 50 \muas $. The first part of Table~1 
is for a magnitude cut-off at $V = 15$ corresponding to the upper 
right panel of Figure~2. The second part of the table is for a 
cutoff at $V = 13$, corresponding to the lower right panel. 

     The effect of the magnitude cut-off on the bias correction 
in column 3 is clear by comparing the first and second parts of 
the table. Note from the second section of the table, and more 
directly from Figure 2, that for distances beyond that where the 
magnitude cutoff line intersects the M = +0.6 line, the bias 
correction of the sample that remains after the magnitude 
restriction {\it changes sign}. This effect, clear from these 
simulations, has been observed in two important papers by 
Oudmaijer, Groenewegen, and Schrijver (1998, 1999) in their 
remarkable demonstrations of the relevant biases empirically. 
They call the reversal of sign of the bias correction at large 
distances (large $ \delta \pi / \pi $) the "incompleteness correction". It 
clearly is a "Malmquist-like" bias, due to a magnitude 
restriction, but here it is not due to an intrinsic dispersion in 
the intrinsic absolute magnitude of the distance indicator 
itself, as in the classical Malmquist correction, but rather is 
due to the dispersion in the {\it inferred} absolute magnitude due to 
the parallax error. Column 6 of Table~1 shows the rms variation 
about the mean bias correction. These are the data needed in   
\S 6.1. 

\subsubsection{The simulations for the density distribution whose 
cumulative increase is as $N(R) \sim R^2$}

     This case is for a true spatial density that decreases 
outward as $R^{-1}$. This requires that the number of stars, $F(R)$, in 
shells of uniform width, $dR$, that are within distance $R$ (i.e. the 
integral function) to increase as $R^2$. Figures 4 and 5 are 
similar to Figures 1 and 2, but for this density distribution that 
decays outward as $R^{-1}$.  
     
     Figure 4 shows the same effect as Figure 1. However, there 
is now less difference between the true distribution (the 
continuous sawtoothed curve) and the ``observer's" distribution 
(the histograms). The effect of the magnitude cutoff is seen well 
in Figures 4 and 5. The magnitude restriction has a stronger 
squelching effect at large distances 
than in Figures 1 and 2, as seen directly from the summary in Table~1.  
   
\subsubsection{The simulations for the density distribution whose cumulative 
increase varies as $N(R) \sim R$}

     Perhaps the most interesting case is that where the spatial 
density decays outward as $R^{-2}$, giving a flat differential 
distribution, i.e. the number of stars at $R$ in interval $dR$, shown 
by the straight line in Fig. 6. This is the $\beta = 2$ case in West 
(1969). Because the number of stars in each shell of width dR is 
the same at all distances, an intuitive guess would be that there 
would be the same number of stars thrown out of the shell at 
small distances as enter the shell from larger distances, and 
therefore there would be no bias. This, however, is not the case. 
Figures 6 and 7 and Table~1 again show the bias, although it is smaller 
than in the previous two cases.      

     The reason for any bias at all is that although the 
distribution of $N(R)dR$ is flat, the distribution in parallax 
space, $G(\pi)d(\pi)$ is not. Equation (1) shows that if $F(R)$ is 
constant, then the distribution in parallax is                          
$G(\pi)d(\pi) = \pi^{-2} d\pi$. This is highly non linear by its $\pi^{-2}$ 
distribution. Hence, a symmetrical error filter in $\pi$, when 
folded into the parallax distribution will still give a different 
value for the mean value of $\pi$ than given by the true  
(errorless) $\pi$ distribution. The effect is identical as that 
demonstrated by Lutz (1978). Of course, the same bias would be obtained 
from the number 
distribution, $F(R) = constant$, by using the {\it asymmetrical} error 
distribution in $R$ that is obtained by transforming the 
{\it symmetrical} error distribution function 
in $\pi$ into the {\it asymmetrical} error 
distribution in $R$.
 
     Figures 6 and 7 show the same statistics for the $F(R) = constant$ 
case [i.e. $N(R) \sim R$] as Figures 1, 2, 4 and 5 
show for the $n = 3$ and $2$ 
cases.      

\subsubsection{Summary of the growth of the bias errors with distance 
on the assumption of a constant parallax error of $50 \muas$ 
at all magnitudes} 

     Figure 8 gives a summary of the results of this subsection 
for the bias magnitude differences from the $M = +0.6$ input value 
for the three cases of $n = 3, 2$ and 1. The data are in Table 1.
The curves would be smoother if there were more statistical samples. 
They appear segmented due to statistical fluctuations in the various bins. 
Figure 8 is, of course, from the unrealistic case of a 
fixed rms parallax error of $50 \muas$ independent of 
apparent magnitude. The {\it FAME} specifications call for a variable 
rms error depending on magnitude. The effect of the increase of 
the rms parallax error with apparent magnitude in the {\it FAME} 
program is in the next section.

\subsubsection{The realistic case of the rms accuracy varying with apparent magnitude}

 The rms parallax error, assumed constant at $50 \muas$ in the previous 
section, does not represent the real expectations for the {\it FAME} experiment. 
Rather, the rms measurement errors with {\it FAME} go from $24 \muas$ at $V=9$, 
through $90 \muas$ at $V=12$ to $443 \muas$ at $V=15$. The run of $\sigma$
with apparent magnitude $m$ is well represented by the relation:
\begin{equation}
\sigma = 10^{(0.21m - 6.547)}
\end{equation}
which is an exponential increase with $m$. Since very few RR Lyraes are 
bright enough to have parallax errors with {\it FAME} as small as $50 \muas$, 
the real case for the bias will be worse. Here we re-run the previous 
simulations, but model the parallax errors according to the relation above.
This provides us with realistic estimates of the bias.

Consider first the case where the spatial density of RR Lyraes is 
constant with distance, for 
which the cumulative number of objects $N$ to distance $R$ increases as $R^3$.
The results are shown in Figures~9 and 10, which are the same as Figs.~2 and 
3 respectively, except we now use the {\it FAME} model for the error estimate 
$\sigma$. The histograms in Fig. 10 show that the results become catastrophic
even at distances of only 1 kpc if there is no magnitude cutoff. 
We summarize the results in part (a) of Table~2, listing the bias 
for each case similar to the listings in Table~1.

Again, there is the apparent oddity in Table~2(a), mentioned in \S 3.3.2, 
where (in each of the 
three  density cases) the bias first 
increases sharply with distance bins, and then decreases as the 
distances get even bigger, eventually changing sign for 
large $\delta \pi / \pi$.
This is because relatively small {\it observed} distances are reported for 
the relatively huge numbers of stars at larger {\it true} distances, due to 
the parallax errors. But as we go further out in observed distance, the 
magnitude cut prevents objects from yet larger distances to be reported 
into these distances -- thus the relative number of pollutants from larger 
distances are kept out, and the bias actually reverses. Again, this is called 
the `incompleteness bias' by Oudmaijer et al. (1998, 1999). 

We have extended the calculations to show 
the simulated values of $\langle M \rangle$ that result from limiting $m$ to  
brighter values, as bright as $m=10$. The results for $m = 13, 12, 11$ and 
$10$ are shown in Fig.~11. 


Fig.~12 is the equivalents of Fig.~11, but for the case 
of $N \sim R^{2}$, or where the density of stars is falling inversely with 
distance ($ \rho \sim R^{-1}$). Predictably, the bias is smaller than for 
the $N \sim R^{3}$ case. Fig.~13 shows the corresponding results for 
the case of $N \sim R$ ($\rho \sim R^{-2}$). The steeper the fall off of density, the 
fewer stars there are at large distance to pollute the sample with stars 
whose measured distances are skewed to shorter distance, and as a result, the 
smaller the bias. Hence, this case shows the least bias, and even at 
the faintest of the four cuts in $m$ ($m < 13$), the bias is quite small.

To compare the realistic {\it FAME} case that has variable parallax errors 
with the magnitude 
cut at $V=13$ for the case of the constant error of $50 \mu as$ in 
section (b) of Table~1 and Figures 2, 5, and 7 we list in section (b)
of Table~2 the data in a similar way as in section (a) for all 3 density 
models. This cut of course greatly reduces the bias effect shown in 
section (a) of the table, and is now quite manageable as a practical matter. 
Although the bias 
is larger for the {\it FAME} model than for the constant error case, 
as is apparent from a comparison of Tables 1 and 2, it is still manageable 
for the $V=13$ cases, shown explicitly in Fig.~14.

The net bias and the rms scatter for the three cases of density and 
four different magnitude cuts are summarized in part (c) of Table~2, listed
in intervals of apparent magnitude.  
We see again, that the bias has a reverse sign, which is direct consequence 
of cutting (severely) in magnitude alone. 
Here the cut in true 
distance (via $m$) means that objects that are truly 
farther than the cut cannot enter the 
sample, thus restricting the number of far away objects that are measured 
as too close, but objects within the sample can be mis-measured as farther 
away than the distance corresponding to the cut, 
thus leading to higher estimates in the mean for $M$. Again, this can 
be appreciated by inspecting Figs. 11, 12 and 13.

\subsubsection{The practical application from these simulations} 

     Can we measure the absolute magnitudes of the RR Lyrae stars 
well enough, as corrected for bias, by using, say, the important 
Layden (1994) sample with its extensive data on metallicity, 
Galactic absorption corrections, and apparent magnitudes? To be 
sure, these data have been variously updated by Layden et al. 
(1996), Layden (1998), Gould and Popowski (1998), and others, and 
will be further improved by the expected photometric data to 
$V = 15$ by the {\it FAME} mission itself. Nevertheless, the 1994 Layden 
table shows a lower limit for the number of available stars. 

     Although the simulations here, plus the many analytical 
solutions in the earlier literature, are useful, we believe they 
are not adequate as a way to eventually be used directly with the 
{\it FAME} catalog because none of them are realistic enough for 
several reasons, such as mentioned at the end of section 2.3 and 
in section 3.1. 

     The simplest of the problems, not answered by the 
simulations nor the analytical solutions, concern the density 
distribution and the sample selection. 
\begin{enumerate}
\item 
What is the appropriate density distribution to use for 
the complete sample?

\item 
Is the density distribution different for different 
RR Lyrae metallicity groups due to the known differences in the 
Galactic spatial distribution as a function of metallicity? 

\item 
The density distribution of the sample that is 
eventually used is likely to differ from the true density 
distribution(s), further complicating the choice of the 
appropriate density distribution. For example, not all {\it FAME} stars 
will have adequate metallicity data. This, therefore, complicates 
the sample selection.

\item   
The actual {\it FAME} parallax errors will be much more 
complicated than simply being a function of apparent magnitude. 
They will also be a non-trival function of ecliptic latitude 
because of orbital restrictions on the space craft (sections 
2.3, 3.1).

\end{enumerate}

     These and other issues can introduce significant differences 
in the bias correction. For example, 
for a cut at $m < 13$, 
the bias ($<M_{inferred} - M_{true}>$ is $-0.14$ mag for the $N \sim R^{3}$ case, 
and only $-0.05$ mag for the $N \sim R$ case.
Hence, the difficulties in any direct application of such simulations done 
here for example, or as calculated previously 
beginning with West (1969) and ending with Hanson (1979) as applied by Reid (1997) to 
real data, is a lack of knowledge of precisely what assumptions should be 
used in the the simulation tables.
Variations as large as 0.2 mag in the differences in the bias corrections
occur depending on the value of $\delta \pi / \pi $ and the assumptions on 
the density distribution; see Table~1 even for the ideal case of constant 
parallax error of $ 50 \muas $. The variations, of course, are larger for 
the realistic case of the {\it FAME} specifications as seen by comparing 
Figures~9-14 with each other as summarized in Table~2. 

Hence, the purpose of all the simulations here has not been to produce 
definitive 
values in Tables 1 and 2 to be used with real data, but rather to demonstrate 
the properties of the bias under a variety of assumptions. Armed now with 
this knowledge from Figures~1-14 and the abbreviated summaries in 
Tables~1 and 2, we discuss in \S 6 a purely empirical method to analyze the 
direct method of this section (not the ``inverse'' method in the next 
section). 

We shall propose in \S 6 to discover empirically the variation of the derived 
absolute magnitudes, star by star, by using binned observational data 
from the {\it FAME} catalog with cuts at various values of $\delta \pi / \pi $ and 
the ``observer's'' distance as well, plus progressive limits in apparent 
magnitude. This empirical approach, guided by the general run of expectations
seen from Figs. 1-14 and Table~2, is the phenomenological approach we expect to 
use with the {\it FAME} data. 

However, in the special case of the RR Lyrae stars where the absolute magnitude at given 
$[Fe/H]$ values is sensibly constant to within narrow limits, we can use 
the very powerful inverse method that ideally would be free from the bias. 
This is the ``maximum-likelihood'' method used by Feast \& Catchpole (1997) 
for the classical Cepheids. It is the method to which we now turn.

\section{A CONDITIONAL METHOD FOR OBTAINING UNBIASSED ABSOLUTE MAGNITUDES}

Assuming as we do for RR Lyraes, that all stars have the same 
absolute magnitude, we effectively know the {\it a priori} distribution of the 
relative distances via the apparent magnitudes, which typically have 
insignificant 
uncertainties (provided the extinction corrections have been done adequately).
In this section we discuss how this assertion provides a 
tractable method for obtaining an estimate of the absolute magnitudes 
that is free of bias by considering the ``inverse'' problem. 

As demonstrated in the previous sections, the bias in the determination 
of $M$ arises because $M$ is not linearly related to the parallax, and 
because the measurement errors, which are expected to have a normal 
distribution in the parallax, propagate to errors with a skewed distribution 
in $M$. Hence, when averages (or other statistics) are computed 
from individual estimates of $M$, they are biassed as a result of the skewed 
distribution in $M$. 

However, if a function $f(M)$ of $M$ can be constructed that 
is linear with the parallax $\pi$, then averages and other statistics on 
$f(M)$ will behave as for a variable with normally distributed errors.
This method was presented by Turon, Lacarrieu \& Cr\'{e}z\'{e} (1977), which is 
paraphrased below.

We re-write the basic relation:

\begin{equation}
 5 \log \pi = M - m - 5
\end{equation}

as 
\begin{equation}
 \pi = 10^{0.2M} / 10^{0.2(m+5)} 
\end{equation}

Let $\pi_{i}$ denote the {\it observed} value of parallax for the $i^{th}$ 
star, 
which differs from the true parallax by an amount $e_{i}$  which is drawn 
from a Gaussian distribution of errors with rms 
$\sigma_{i} \equiv \delta \pi$. Thus, for the 
$i^{th}$ star,
\begin{equation}
 Y \equiv f(M) = 10^{0.2M} =  \alpha_{i} (\pi_{i} - e_{i}) 
\end{equation}
where $Y$ represents the {\it true} value of $f(M)$, 
$\alpha_{i} = 10^{0.2(m_{i} + 5)}$ and $m_{i}$ is the apparent 
magnitude of the $i^{th}$ object, corrected for extinction.
Given a value for Y and for $m_{i}$, one can then write the probability 
$P(\pi_{i})$ of
observing a parallax $\pi_{i}$ as: 
\begin{equation}
 P(\pi_{i}) = \frac{1.}{\sqrt{2\Pi} \, \sigma_{i}} \exp \left( - \frac{(Y - \pi_{i}\alpha_{i})^2} { 2 \sigma_{i}^2 \alpha_{i}^2} \right)
\end{equation}
The likelihood ${\cal L}$ of observing a set of specific values of $\pi_{i}$ for 
$i = 1$ to $n$ is the product of the probabilities $P_{i}$ above::
\begin{equation}
{\cal L} = {\cal A} \exp \left( - \sum_{i=1}^{n} \frac{(Y - \pi_{i} \alpha_{i})^2} {(2 \sigma_{i}^2 \alpha_{i}^2)} \right)
\end{equation}
where ${\cal A}$ is the normalization term. 
The maximum likelihood value for $Y$ is obtained when we set
\begin{equation}
\frac {\partial {\cal L}} {\partial Y} = 0
\end{equation}
which implies 
\begin{equation}
<Y> = f(M)_{maxlikely} = \frac{ {\displaystyle \sum_{1=1}^{n}} \frac{ \pi_{i} \alpha_{i} } {\sigma_{i}^2 \alpha_{i}^2} } { {\displaystyle \sum_{i=1}^{n}} \frac{ 1 }{\sigma_{i}^2 \alpha_{i}^2} } 
\end{equation}
This estimate is linear in the measured values of $\pi_{i}$, and so if 
the error in measuring the $\pi_{i}'s$ is unbiassed, $f(M)$ will also be 
unbiassed. The likelihood of obtaining a value $<Y>$ for the optimally 
weighted mean that differs from the true value of $Y$ by an amount 
$\delta Y$ is  again a Gaussian distribution:
\begin{equation}
{\cal L} = {\cal A} \exp \left( - \left( \frac{(\delta Y)^2}{2} \right) \sum_{i=1}^{n} \frac{1} {(\sigma_{i}^2 \alpha_{i}^2)} \right)
\end{equation}
This shows again, that the calculated mean $<Y>$ is normally distributed 
about the true value $Y$, with an effective rms (or $\sigma$) error 
$E(f(M))$ given by 
\begin{equation}
E(f(M)) = \left( \sum_{i=1}^{n} \frac{ 1 }{\sigma_{i}^2 \alpha_{i}^2} \right)^{-1/2}
\end{equation}

This asymptotic expression for $f(M)$  was given by Turon Lacarrieu \& 
Cr\'{e}z\'{e} (1977), who indicate that it was used earlier by Roman (1952) and by 
Ljunggren and Oja (1965). More recently, it has been used by Feast \& Catchpole
(1997) in their work on Cepheid parallaxes with the {\it Hipparcos} satellite,
and also by Koen \& Laney (1998). 
The method of weighting is such that objects with large errors are correctly
down-weighted, and one does not need to worry about trimming an observational 
sample, {\it as long as good estimates for the parallax errors exist}.

The most probable value and upper and lower $n- \sigma$ limits for $M$ 
are given by:
\begin{eqnarray}
& & M(best) = 5 \log f(M)  \\
& & M(+n \sigma) = 5 \log (f(M) + nE(f(M))  \\
& & M(-n \sigma) = 5 \log (f(M) - nE(f(M)) 
\end{eqnarray}

It is worth understanding this procedure in physical terms. As we have seen 
in the previous sections, the bias has two components: 1) the asymmetry in 
the upper and lower bounds of errors in $M$ because $M$ depends 
logarithmically on the parallax, for which the errors are normally 
distributed and hence symmetric, and 2) due to the relative object densities 
at different distances. The procedure above calculates the means and errors 
in the space of $f(M)$ which is linear with respect to the parallax, thereby 
taking care of concern (1) above. In addition, it utilizes the assertion that 
$M$ is constant for all objects in the sample, and then uses the apparent 
magnitude as a relative indicator of distance, and through the $\alpha$ 
term, compensates the weights to correct for the space density bias. 
If $M$ is not constant, but if its variation is 
predictable via some other parameter 
(period, metallicity, etc; see Feast 1999), one can 
still apply the above method to solve for the zero-point of the 
parametric relation.
An intrinsic spread $\sigma_{M}$ of the absolute magnitude $M$, if comparable,
or larger than the individual fractional parallax errors, will contribute
not only to additional uncertainty in determining $M$, but will also 
create a Malmquist bias for any real sample. We simulate the magnitude of these
effects for {\it FAME} parallaxes of a representative sample of RR Lyrae 
stars in \S 6.2, and find them to be insignificant for a realistic spread
$\sigma_{M}$.
 

We have tested the efficacy of the above method on the simulated samples 
discussed in \S 3. We have verified, by generating multiple runs with 
independently random generating the errors in parallax, that irrespective 
of how the samples are trimmed, the method above produces values 
of $\langle M \rangle$ that are in agreement with the assumed input value of $0.60$ mag
to within 2-$\sigma$ of the errors computed by equations (14) and (15).

\section{THE DISTRIBUTION OF RR LYRAE STARS IN THE LOCAL NEIGHBORHOOD BINNED BY DISTANCE AND METALLICITY}

\subsection{The available RR Lyrae stars within 1600 parsecs that have metallicities and absorption values from Layden's list}

     The discovery of all the RR Lyrae stars within say 3000 
parsecs of the sun is not complete (e.g. Layden 1994, Fig. 10). 
Even more incomplete is the determination of metallicity for the 
variables that are known within this distance. Nevertheless, 
important survey lists do exist from which we can estimate the 
level of success that can be achieved by {\it FAME} in the calibration 
of $M_{V}(RR) = f([Fe/H])$ using even the present incomplete data 
set.
     An important modern summary is that of Layden (1994). He lists 
most of the entries of RRab variables from the 4th edition of the 
General Catalog of Variable Stars (Kholopov 1985) that are at 
Galactic latitude greater than $10^{o}$. He has also re-determined 
metallicities for all his listed variables on the scale of Zinn 
and West (1984). His list contains a combination of new 
observations of metallicity by him with a homogenization of 
other data in the literature.  
     Layden's compilation contains 303 RRab Lyrae stars. Among 
other things, he lists (1) his newly derived metallicities, (2) 
apparent $V$ magnitudes at mean light, most of which are from 
photo-electric photometry, (3) derived visual Galactic absorption, 
(4) Layden's distances\footnote{Layden's distance scale is compressed (smaller) compared with 
the scale based on the calibration using the 
Oosterhoff-Arp-Preston period-metallicity effect and the pulsation equation 
that requires $M_{V}(RR) = 0.30[Fe/H] + 0.94$ (Sandage 1993a). This 
calibration gives larger distances than the Layden scale by 
factors that range between 1.05 and 1.23, depending on [Fe/H]).}
are based on $<M_{V}> = 0.15[Fe/H] + 1.01$ and the 
corrected $<V>$ apparent magnitudes. 

     We have binned Layden's list into three distance ranges in 
Table 3 in order to determine how many RR Lyrae stars there will 
be for the {\it FAME} calibration in each distance and metallicity 
range. The tabulation of the Layden list for RR Lyrae stars that 
are expected to have parallax errors of less than $(\delta \pi/ \pi)$ 
of 0.08 in the first two parts of Table 3 and less than 0.12 in 
the third part of the table, according to the adopted Layden 
distance scale. 
     The table is divided into three sections separated by
distances of (a) less than 1 Kpc, (b) between 1.01 and 
1.25 Kpc, and (c) 1.26 and 1.40 kpc on Layden's scale. 
These 
distances are smaller than what we believe to be the correct 
calibration required by the Oosterhoff-Arp-Preston period 
metallicity effect. Hence, although the 
subsamples in Table 3 have been selected by the Layden tabulated 
distances, the distances in column 7 of Table 3 have 
been calculated using the larger distance scale based on the just 
mentioned Oosterhoff effect. 
     Hence, the expected parallaxes in column 8 of Table 3 are 
based on the largest distances encountered in the current 
literature on the calibration of $M_{V}(RR)$, and as such, are the 
most pessimistic concerning the capabilities of the {\it FAME} 
calibration. Said differently, if the Table 3 data show that the 
{\it FAME} mission can accomplish the calibration, we have used the 
most stringent data here to prove it.         

Column 2 of Table 3  shows the metallicity listed by Layden 
(1994), based on his own re-calibrations. Column (3) is the 
assumed absolute magnitude according to the adopted calibration 
of $M_{V}(RR) = 0.30([Fe/H]) + 0.94$ (Sandage 1993a) that differs 
from Layden's assumption. Columns (4) and (5) are the adopted 
mean $V$ magnitude and the Galactic absorption according to 
Layden. Column (6) is the resulting absorption-free mean $V$ 
magnitude found by combining columns (4) and (5). 
Column (7) is column (6) minus column (3). Column (8) 
is the predicted parallax from the distance implied in column 
(7). Column (9) is the expected rms parallax error taken from the 
{\it FAME} specifications. Column (10) is the ratio of columns (8) and 
(9).

     Table 3 lists the principal RR Lyrae stars of highest weight 
(the nearest) that will be available for the {\it FAME} RR Lyrae 
absolute magnitude calibration. The {\it FAME} database will, of 
course, contain many more such stars that are fainter, at least to 
$V = 14.5$, and these will be highly useful in strengthening the 
empirical bias correction for the direct method. However, it is a 
list of the nearest stars, similar to these in Table~3, updated 
of course, that will carry the bulk of the weight for the 
absolute magnitude and that will have the smallest bias 
correction. If the number of stars in this nearby list is 
inadequate to beat down the rms errors about the mean bias values 
in Table~2 (columns 6, 7, and 8 for stars closer than D = 1500 pc) 
to adequately small values, then the calibration via the 
direct method will be compromised. We address this question of 
the number of available stars in the next section.

\section{THE FEASIBILITY OF USING {\it FAME} PARALLAXES}

\subsection{Using the direct method that requires Lutz-Kelker bias corrections}

When the {\it FAME} catalog of measured parallaxes with their error
estimates becomes available, there is no doubt that the RR Lyraes calibrations
will be analyzed by both the direct method of \S3.3 and by the ``inverse'', conditional 
method of \S 4. For the direct method, given the rms dispersions listed in 
Table~2 for the {\it FAME} error model (and shown in Figures 9-14), we inquire in this 
section if there will be a large enough sample of RR Lyrae stars, as listed in Table~3,
when binned by $[Fe/H]$, and, in addition, either by apparent magnitude or 
``observer's'' distance (i.e. measured parallax) to beat down the rms errors in Table 2
to sufficiently small values. 

To answer this question of population statistics we have binned Table~3 by 
metallicity and apparent magnitude. The apparent magnitude binning (if all 
variables of a particular metallicity have identical absolute magnitudes) will, 
of course, be a binning by {\it true} distance, whereas a binning by the catalog
values of $(\delta \pi / \pi)$ will be by the ``observer's'' distance. The two 
distances will be nearly the same for small enough ($\delta \pi /\pi)$ values, but 
will diverge progressively as the relative error becomes larger. Because of this complication,
and because of the  questions set out at the end of \S 3.3.7 it will be easiest to 
proceed entirely empirically to assess the errors by analysing the derived mean RR 
Lyrae absolute magnitudes as functions of the measured $(\delta \pi / \pi)$ values
which, of course will also be a strong function of apparent magnitude. Are there 
enough stars with small $(\delta \pi / \pi)$ values in each metallicity bin to beat down the 
rms values given in Table~2?

Binning by the listed $(\delta \pi / \pi)$ values in Table~3 for three bins of metallicity 
shows 22 RR Lyrae stars with $[Fe/H]$ between $+0.07$ and $-0.99$, 29 with $[Fe/H]$ 
between $-1.00$ and $-1.49$, and 32 stars with $[Fe/H] < -1.50$ for $(\delta \pi / \pi)$ 
values less than 0.14. It can be shown from Table~3 that this upper limit of $(\delta \pi / \pi)$
corresponds to a limit of $V=12$, which is even more advantageous than the data in Table~3 
for $V=13$. The $(\delta \pi / \pi)$ limit corresponding to a limiting magnitude $V = 13$ 
is about 0.20; Table~3 could be expanded by more than a factor of two from Layden's (1994) 
master list if we accept this larger value of the relative parallax error. 

We will of course analyze the complete {\it FAME} catalog to values of $(\delta \pi / \pi)$
this large and larger in order to see well the bias which, according to Figure~8, will 
be primarily a function of the relative error, as all simulations, beginning with 
West, have shown. 

Dividing the rms values in Table~2 in the entries with 
$ \delta \pi / \pi $ by the square root of the number of stars 
in each of the three metallicity bins listed above, shows that the
{\it FAME} error budget is entirely adequate to determine the constants $a$ and $b$
in the relation $M(RR) = a [Fe/H] + b $
to an accuracy of a few hundredths of a magnitude for $b$ and to within 
$\sim 15\%$ for $a$.

     For example, using the magnitude cut between $ m = 11$ and 
$ m = 12 $ (to imitate the magnitude range in Table~3) over the entire 
distance range (part c of Table~2) the relevant rms values are 
$ \sim 0.15 $ mag. (See also Figures 11-13 for the smallness of the 
bias error and its rms values with these magnitude restrictions). 
Hence, with this value of the rms about the mean bias value, the 
accuracies with which we shall be able to determine the mean bias 
value for each of the metallicity bins, even if only the Table~3 
list were to be used out of the entire {\it FAME} database, are : 0.032 
mag for $ <{\rm  [Fe/H]}> =  -0.53$ with the 22 stars in Table~3, 0.028 mag 
for $ <{\rm [Fe/H]}> = -1.25 $ with the 29 stars in Table~3, and 0.027 mag 
for $ <{\rm [Fe/H]}> = -2.0 $ with the 32 stars in Table~3. 
These rms errors on the accuracy with which the bias 
correction can be determined will of course be smaller, the larger 
the number of stars used for the determination. Table~3 
is indeed the minimum list. It can be expanded by at least a 
factor of two by increasing the distance restriction imposed 
arbitrarily there to at least 2000 kpc, at which distance the 
$ \delta \pi / \pi$ will still be less than 0.20 for the {\it FAME} 
data if the proposed {\it FAME}  accuracies are still those of equation (4). 

     Hence there is every prospect of determining an RR Lyrae 
absolute magnitude calibration to considerably better than  0.1 
mag even when the data are analyzed separately by metallicity 
range. Indeed, even using the small sample of Table~3 with the 
errors listed above, the slope coefficient to the metallicity 
dependence, $ dV / d{\rm [Fe/H]} $, can be determined to within 
$ \pm 0.04$  about a value of 0.30, or to within 13\%. 
This should bring 
to a close the present controversy over a difference in this 
slope, presently standing at a factor of two between the long and 
the short distance-scale groups. Of course, even greater accuracy 
will ensue using the much larger total {\it FAME} database, even with 
the increased bias corrections which can be controlled 
empirically. Similar accuracies are expected using the inverse 
maximum likelihood method as we show in the next section.

\subsection{Using the maximum likelihood method of \S 4}

To test whether the parallaxes from {\it FAME} are good enough to obtain 
the absolute magnitudes for the RR Lyrae stars in the Layden sample
(and sub-samples thereof for testing metallicity dependence), we need 
to see if equation (13) gives sufficiently small values for this 
sample, and for the desired sub-samples.

The procedure is straightforward:
\begin{enumerate}
\item
Beginning with the extinction corrected magnitudes (column 6 of Table~3), 
compute the rms parallax error $\sigma$ for each object according to 
equation (4). Also calculate values of $\alpha = 10^{0.2(m+5)}$.
\item
Compute $E(f(M))$ from equation (13) for the desired sub-sample of objects.
\item
Assume a value for the absolute magnitude, say +0.60.
Using the apparent magnitudes (extinction corrected), calculate their true 
parallaxes. Using a random draw from a Gaussian distribution with the 
pertinent $\sigma$ for each individual object, assign a parallax error, and 
add it to the true parallax to obtain a simulated observed parallax.
\item
Using the observed parallax and the $\alpha$ and $\sigma$ for each object in 
any selected sub-sample of object from Layden's list, calculate the best 
value for the observed absolute magnitude for the simulation using equations
(11) and (14).
\item 
Calculate the upper and lower error bounds using equations (15) and (16)

\end{enumerate}

The results of such a round of simulations using Table~3 directly are shown 
in Table~4. The 
$1 \sigma$ bounds do not always bracket the assumed value of $M = +0.60$, 
but the $2 \sigma$ bounds do. Clearly the errors are small, and the 
results would provide very stringent constraints on the intrinsic 
brightness of the RR Lyrae stars. The sub-sample of stars with 
$[Fe/H] < -1.5$ and $[Fe/H] > -1.0$ differ in metallicity by approximately 
$\Delta({\rm [Fe/H]})$ of unity. 


To estimate the effect of an intrinsic spread $\sigma_{M}$ in the absolute 
magnitude $M$, the following simulations were done with the Layden sample.
Drawing from a normal distribution about 
an assumed value of $\sigma_{M}$, the 82 stars in the sample were assigned 
individual random deviations from $M = +0.60$. Random errors in parallax 
were also assigned from the {\it FAME} model (equation 4). The maximum 
likelihood method was then used to deduce an `observed' value for $\langle M \rangle$.
The simulation was run 100 times for a given value of $\sigma_{M}$, 
with independent random values generated for the deviation in $M$ and 
error in $\pi$ for each star in each of the runs. Thus we obtain 100 
independently sampled values for the `observed' $\langle M \rangle$. The mean 
and rms values of this recovered value of $\langle M \rangle$ for various values of 
$\sigma_{M}$ are shown in Table~5. As expected, the rms uncertainty 
in determining $\langle M \rangle$ increases with $\sigma_{M}$. This rms uncertainty 
is the realistic accuracy with which we can expect to estimate $\langle M \rangle$.
 
Since there were 100 trials, we can track the systematic error due to biases
to a precision given by the error 
in the mean of simulated $\langle M \rangle$, i.e.  $ \sim 1/10 $ the rms in $\langle M \rangle$.
This means that for the values shown in Table~5, the values of $\langle M \rangle$ are 
estimated with uncertainties of a few hundredths of a magnitude at worst. 
The run of $\langle M \rangle$ versus $\sigma_{M}$ in Table~5 is thus an estimate 
of the Malmquist bias as a function of $\sigma_{M}$. 
The second set of rows in Table~5 show the same results for one of 
the metallicity sub-groups in the Layden sample. The bias and scatter 
are a little larger, as expected for the fewer number (23) of stars, but are 
not radically worse than for the full list.

The bias depends 
not only on $\sigma_{M}$, but also on the spatial distribution of objects
in the sample, and the rules of inclusion in the sample. For the final 
list of objects with {it FAME} parallaxes that is actually used, the 
simulation as above will need to be re-run, but the Layden sample 
provides a good estimate of what to expect. 

The intrinsic spread $\sigma_{M}$ has been measured in globular clusters.
Sandage (1990b) finds the standard deviation to vary from 0.06 mag to 0.15 mag
in the ${\rm [Fe/H]}$ range $-2.2$ to $-0.7$, with the trend that there 
is larger spread at higher metallicities. 
In conjunction with Table~5, this 
means that applying the maximum likelihood method 
to {\it FAME} parallaxes for even a minimal sample of RR Lyrae stars 
like the Layden list will produce estimates of the mean absolute magnitudes 
broken down by broad metallicity bins to uncertainties within $\sim 0.05$ mag.

We learned, after this paper was written, that `de-scoping' the {\it FAME} project
in response to budget issues will result in parallax errors that are larger 
than the earlier estimates. Specifically, the accuracy is now estimated to be
$50 \muas$ at $V=9$, and $700 \muas$ at $V=15$, where the earlier numbers 
were $ 24 \muas $ and $ 443 \muas $ respectively. 
The discussions in this paper remain qualitatively unchanged, though of course 
the bias effects shown for the `{\it FAME}' model will be worse for the experiment
as now modified. Nevertheless, the specific problem of RR Lyrae 
absolute magnitudes and dependence on metallicity remains tractable even with 
these cuts in accuracy.

  The first author is indebted to Andrew Gould for an enlightening 
conversation (during a meeting) concerning the power of the 
maximum likelihood method of 
analysis where the direct Eddington/Lutz-Kelker bias is moot, although 
Malmquist bias remains.  We are also greatly 
indebted to Gould for his detailed, thorough and thoughtful 
refereeing of 
the first draft of this paper. We agree with, and have incorporated all 
of his technical suggestions in the final manuscript. We thank George 
Wallerstein 
for alerting us to the work of Oudmaijer et al., who have approached 
the bias problem empirically by using Spaenhauer diagrams. We are also grateful to 
Gustav Tammann and Pekka Teerikorpi for their reading and comments on 
the manuscript.

\newpage
%
%

\newpage
%
%

\figcaption[]{Comparison of the true distribution of stars with 
distance (the continuous saw-toothed curve) with the apparent 
distribution (the histogram with bin size of 50 parsecs) in an 
``observers catalog" made from the true distribution by using 
Gaussian parallax errors for each star with an rms Gaussian width 
of $50 \muas $. The true cumulative number of stars within 
distance R is assumed to vary with distance as $R^{3}$, given by a 
constant density with distance. Four comparisons are shown for 
the cases of no restrictions on apparent magnitude in the 
``observer's catalogue'', and for the three magnitude cut-offs of V 
= 15, 14, and 13. The simulation is made using $3 \times 10^{6}$ stars 
distributed uniformly in a volume of 50,000 parsecs.} 

\figcaption[]{ Distribution of inferred absolute magnitudes from the 
``observer's'' catalog as a function of distance for stars  which 
all have a true absolute magnitude of +0.6 (at the white stripe) 
using the histogram distributions in Figure 1. The three panels 
with magnitude cuts show the effectiveness of excluding stars 
that enter the distance ranges here from large true distances 
because the error in distance due to a parallax error of $d\pi$ 
increases as $dR = R^{2} d(\pi)$, and is very large at large distances.}

\figcaption[]{Distribution of the magnitude errors due to bias as a 
function of the ``observer's'' distance from 2000 parsecs to 4000 parsecs, 
in distance intervals of 
500 parsecs, from the distribution in Fig. 1 using an apparent 
magnitude cut-off at V = 15. The mean absolute magnitude, the rms 
of the magnitude distribution, and the number of stars dN making 
the histogram are marked on the diagrams.}

\figcaption[]{Same as Fig. 1 but for a cumulative count number, $N(R)$, 
that varies as $R^{2}$ in the true catalog. This corresponds to a 
spatial density that decays at the rate of $R^{-1}$.}

\figcaption[]{Same as Fig. 2 but for the $N(R) \sim R^{2}$ case of Fig. 4. The 
number of stars in this distance range to 5000 parsecs is larger 
than in Fig. 1 because fewer of the total number of $3 \times 10^{6}$ stars 
are at larger distances due to the different assumed distribution 
of the stars.}     

\figcaption[]{Same as Fig. 1. but for the case where the number of 
stars within a distance R increases only as $N(R) \sim  R$. This 
gives a constant number of stars in each shell of width $R$ at each 
distance, i.e. the differential count $dN(R)$ is independent of 
distance as shown by the level line in each panel. However, the 
differential distribution in parallax is not flat, decreasing 
with increasing parallax as $\pi^{-2}$. Hence, with a symmetrical 
parallax error distribution there still is a bias effect because 
the histogram distribution in each panel is above the level line 
except near the cut-off region due to the magnitude cut.}

\figcaption[]{Same as Fig. 2. but for the $N(R) \sim R$ case of Fig. 6. 
Again, the density of stars is greater than in Figs. 2 and 5 
because of the difference in the assumed density distributions.}   

\figcaption[]{The variation of the bias with the relative parallax 
error, $\delta \pi / \pi$ at the midpoint of each distance interval 
for the simulations where the parallax rms error is $50 \muas$ 
at every distance. The curves show the bias for the three 
values of the density distribution with $n = 3, 2$, and $1$, where 
the number of stars within distance $R$ varies as $R^{n}$.}

\figcaption[]{The $N(R) \sim R^{3}$ case for the realistic {\it FAME}
model of the parallax errors as a function of apparent magnitude according to 
equation (4). Three magnitude cutoffs are shown for 15, 14, and 13 mag. 
These represent cuts in the {\it true} distances of $m-M = 14.4,~ 13.4$ and 
$12.4$  ($7.6, ~  4.8,$ and $3.0$ kpc) respectively. Note that these 
distances are where the upper envelope line meets the stripe for the 
assumed absolute magnitude $M=+0.6$. All points above the stripe are 
thrown into nearer distances from more distant points. All stars below 
the stripe are thrown from closer distances to larger ones.}

\figcaption[]{Histograms of the rms deviations for various ``observer's''
distance intervals with a magnitude cut at $V=15$ for the $N \sim R^{3}$
case.}

\figcaption[]{The {\it FAME} error model for $\sigma$ with magnitude cuts 
at $V=13, ~ 12, ~ 11,$ and $10$, continuing Fig. 9.}

\figcaption[]{The {\it FAME} error model for $N(R) \sim R^{2}$ case with 
magnitude cuts at $V=13, ~ 12, ~ 11,$ and $10$ mag.}

\figcaption[]{The {\it FAME} error model for $N(R) \sim R$ case with 
magnitude cuts at $V=13, ~ 12, ~ 11,$ and $10$ mag.}

\figcaption[]{Comparison of the $\sigma = 50 \mu as$ constant error model
with the {\it FAME} error model for the $N(R) \sim R^{2}$ and $N(R) \sim  R$ 
cases with a magnitude cut at $V=13$. The panels are abstracted from 
Figs. 5, 7, 11, and 13.}
\clearpage
\setcounter{table}{0}
%
\begin{deluxetable}{cccccccc}
\tablecaption{Simulated Bias Corrections for Three Density Distributions 
using an RMS Parallax Error of $50 \muas$ that is Constant with Distance}
\tablewidth{0pt}
\tablehead{
\multicolumn{1}{c}{Distance}  & \multicolumn{1}{c}{$d \pi / \pi$}  &  
\multicolumn{3}{c}{Bias $\Delta M$} & \multicolumn{3}{c}{RMS of the Bias} \\
pc  & mid point &  n=3 & n=2  & n=1   &  n=3  & n=2 &  n=1  \\
 &  &  \multicolumn{3}{c}{mag} & \multicolumn{3}{c}{mag}  \\
   (1)   &    (2)  &    (3)  &  (4) &   (5)  &    (6) &   (7)  &  (8) \\
}
\startdata
\multicolumn{8}{c}{ (a)  MAGNITUDE CUT AT V = 15 ($M_{true} = +0.6$) } \\
 & \\
  0 - 500 &  0.013 &   0.000 & 0.001 & 0.001 &   --- &   0.047 & 0.031 \\
  0 -1000  &  0.025  &   0.020  &  0.010  &  0.005  &   0.079  &  0.078  &  0.064 \\
1000-1500  &  0.063  &   0.047  &  0.035  &  0.022  &   0.176  &  0.141  &  0.141 \\
1500-2000  &  0.087  &   0.110  &  0.058  &  0.043  &   0.222  &  0.197  &  0.195 \\
2000-2500  &  0.113  &   0.155  &  0.113  &  0.076  &   0.269  &  0.268  &  0.260 \\
2500-3000  &  0.138  &   0.242  &  0.164  &  0.114  &   0.377  &  0.337  &  0.329 \\
3000-3500  &  0.163  &   0.320  &  0.253  &  0.164  &   0.429  &  0.421  &  0.399 \\
3500-4000  &  0.188  &   0.402  &  0.306  &  0.207  &   0.461  &  0.467  &  0.452 \\
 \\
\multicolumn{8}{c}{ (b)  MAGNITUDE CUT AT V = 13 ($M_{true} = +0.6$) } \\
 & \\
  0 - 500  &  0.013  &   0.001  &  0.005  &  0.001  &   ---    &  0.040  &  0.032 \\
  0 -1000  &  0.025  &   0.010  &  0.009  &  0.004  &   0.071  &  0.078  &  0.063 \\
1000-1500  &  0.063  &   0.016  &  0.034  &  0.022  &   0.160  &  0.147  &  0.140 \\
1500-2000  &  0.087  &   0.050  &  0.060  &  0.044  &   0.215  &  0.205  &  0.197 \\ 
2000-2500  &  0.113  &   0.080  &  0.076  &  0.050  &   0.225  &  0.228  &  0.235 \\
2500-3000  &  0.138  &  -0.042  & -0.052  & -0.072  &   0.187  &  0.197  &  0.203 \\
3000-3500  &  0.163  &  -0.197  & -0.184 &  -0.212  &   0.172  &  0.150  &  0.164 \\
3500-4000  &  0.188  &  -0.550  & -0.559 &  -0.567  &   0.116  &  0.129  &  0.133 \\
\enddata
\end{deluxetable}
\newpage
\clearpage
%
\begin{deluxetable}{cccccccc}
\tablecaption{Simulated Bias Corrections for Three Density Distributions
using the {\it FAME} Model for RMS Parallax Error}
\footnotesize 
\tablewidth{0pt}
\tablehead{
\multicolumn{1}{c}{Distance}  & \multicolumn{1}{c}{$d \pi / \pi$}  &  
\multicolumn{3}{c}{Bias $\Delta M$} & \multicolumn{3}{c}{RMS of the Bias} \\
pc  & mid point &  n=3 & n=2  & n=1   &  n=3  & n=2 &  n=1  \\
 &  &  \multicolumn{3}{c}{mag} & \multicolumn{3}{c}{mag}  \\
   (1)   &    (2)  &    (3)  &  (4) &   (5)  &    (6) &   (7)  &  (8) \\
}
\startdata
\multicolumn{8}{c}{\bf Assume $M_{true} = +0.60$} \\
\multicolumn{8}{c}{ a) by inferred Distance  and using a magnitude cut at  $V = 15$ } \\
  0 - 500  &  0.003  &   0.000  &  0.001  &  0.000  &   0.004  &  0.015  &  0.011 \\
  0 -1000  &  0.012  &   2.800  &  0.771  &  0.077  &   2.140  &  1.866  &  0.558 \\
1000-1500  &  0.075  &   3.020  &  2.093  &  0.901  &   1.148  &  1.626  &  1.415 \\
1500-2000  &  0.150  &   2.242  &  1.710  &  1.096  &   0.940  &  1.148  &  1.141 \\
2000-2500  &  0.252  &   1.645  &  1.342  &  0.932  &   0.823  &  0.932  &  0.943 \\
2500-3000  &  0.380  &   1.295  &  1.026  &  0.715  &   0.743  &  0.819  &  0.841 \\
3000-3500  &  0.535  &   0.970  &  0.784  &  0.483  &   0.683  &  0.760  &  0.783 \\
3500-4000  &  0.717  &   0.752  &  0.502  &  0.285  &   0.653  &  0.722  &  0.741 \\
\\
\multicolumn{8}{c}{ b)  by inferred Distance and using a magnitude cut at  $V = 13$ } \\
  0 - 500  &  0.003  &   0.000  &  -0.004  &  0.000  &   0.021  &  0.014  &  0.010 \\
  0 -1000  &  0.012  &   -0.004  &  -0.011  &  0.005  &   0.596  &  0.060  &  0.049 \\ 
1000-1500  &  0.075  &   0.235  &  0.187  &  0.111  &   0.436  &  0.335  &  0.278 \\
1500-2000  &  0.150  &   0.418  &  0.258  &  0.243  &   0.437  &  0.448  &  0.405 \\ 
2000-2500  &  0.252  &   0.143  &  0.147  &  0.055  &   0.358  &  0.323  &  0.353 \\
2500-3000  &  0.380  &   -0.172  &  -0.190  &  -0.237  &   0.283  &  0.291  &  0.307 \\
3000-3500  &  0.535  &   -0.425  &  -0.464  &  -0.505  &   0.239  &  0.312  &  0.240 \\
3500-4000  &  0.717  &   -0.455  &  -0.379  &  -0.449  &   0.222  &  0.170  &  0.234 \\
 \\
\multicolumn{8}{c}{ c) using a magnitude cut alone } \\
 \\
\multicolumn{2}{l}{Restriction by mag} &
\multicolumn{6}{c}{--------------------------------------------------} \\
$ m < 10.00 $  &     &  0.003  &  0.000  &  0.000  &    0.033  &  0.037  &  0.026 \\
$ m < 11.00 $  &     &   0.001 & -0.002 & -0.001  &    0.106  &  0.088  &  0.067 \\
$ m < 12.00 $  &     & -0.057 & -0.012 & -0.007  &    0.296  &  0.230  &  0.176 \\
$ m < 13.00 $  &     &  -0.137 & -0.074 &  -0.049  &    0.729  &  0.683  &  0.551 \\
\\
\enddata
\end{deluxetable}

\newpage
\clearpage
%
\begin{deluxetable}{lccrcrrrcc}
\tablecaption{}
\footnotesize
\tablewidth{0pt}
\tablehead{
STAR &  [Fe/H] &  $M_{V}$ &
\multicolumn{1}{c} {$<V>$} &  
$A_{V}$ &  
\multicolumn{1}{c} {$<V>^{0}$} & 
\multicolumn{1}{c} {$(m - M)_{s}$} & 
\multicolumn{1}{c} { $\pi$ } & 
$\delta \pi$ & $\delta \pi / \pi$ \\
\multicolumn{1}{c} {(1)}  & 
\multicolumn{1}{c} {(2)}  &  
\multicolumn{1}{c} {(3)} & 
\multicolumn{1}{c} {(4)} & 
\multicolumn{1}{c} {(5)}  &
\multicolumn{1}{c} {(6)}  & 
\multicolumn{1}{c} {(7)}  & 
\multicolumn{1}{c} {(8)}  & 
\multicolumn{1}{c} {(9)} & 
\multicolumn{1}{c} {(10)} \\
 & & & & & & & 
\multicolumn{1}{c} { $\muas$ } & 
\multicolumn{1}{c} { $\muas$ } &   \\
}
\startdata
\multicolumn{10}{c}{ D(Layden) $<$ 1 kpc} \\
 & \\
SW And &  -0.38 & 0.83 &  9.69 & 0.14 &  9.55 &  8.72 & 1802 & 30 & 0.017 \\
XX And &  -2.01 & 0.34 & 10.68 & 0.13 & 10.55 & 10.21 &  908 & 50 & 0.055 \\
WY Ant &  -1.66 & 0.44 & 10.82 & 0.18 & 10.64 & 10.20 &  912 & 50 & 0.055 \\
V341 Aql & -1.37 & 0.53 & 10.87 & 0.31 & 10.56 & 10.03 &  986 & 50 & 0.051 \\
X Ari  &  -2.40 & 0.22 &  9.54 & 0.50 &  9.04 &  8.82 & 1721 & 30 & 0.012  \\
 \\
RS Boo &  -0.32 & 0.84 & 10.35 & 0.00 & 10.35 &  9.51 & 1253 & 40 & 0.032   \\
W CVn &   -1.21 & 0.58 & 10.57 & 0.00 & 10.57 &  9.99 & 1005 & 47 & 0.047 \\
RR Cet &  -1.52 & 0.48 &  9.69 & 0.02 &  9.67 &  9.19 & 1451 & 32 & 0.022 \\
RV Cet &  -1.32 & 0.54 & 10.84 & 0.02 & 10.82 & 10.28 &  879 & 50 & 0.057 \\
XZ Cet &  -2.27 & 0.26 &  9.50 & 0.00 &  9.50 &  9.24 & 1419 & 30 & 0.021 \\
 \\
V413 CrA & -1.21 & 0.58 & 10.62 & 0.32 & 10.30 &  9.72 & 1138 & 45 & 0.039     \\
XZ Cyg &  -1.52 & 0.48 &  9.62 & 0.33 &  9.29 &  8.81 & 1730 & 31 & 0.012 \\
DM Cyg &  -0.14 & 0.90 & 11.55 & 0.69 & 10.86 &  9.96 & 1019 & 73 & 0.072 \\
DX Del &  -0.56 & 0.77 &  9.94 & 0.32 &  9.62 &  8.85 & 1698 & 35 & 0.021 \\
SU Dra &  -1.74 & 0.42 &  9.81 & 0.00 &  9.81 &  9.39 & 1324 & 33 & 0.025 \\
 \\
SW Dra &  -1.24 & 0.57 & 10.52 & 0.04 & 10.48 &  9.91 & 1042 & 45 & 0.043  \\
XZ Dra &  -0.87 & 0.68 & 10.19 & 0.22 &  9.97 &  9.29 & 1387 & 38 & 0.028 \\
RX Eri &  -1.30 & 0.55 &  9.71 & 0.08 &  9.63 &  9.08 & 1521 & 32 & 0.021 \\
SV Eri &  -2.04 & 0.33 &  9.92 & 0.19 &  9.73 &  9.40 & 1318 & 35 & 0.026 \\
SS For &  -1.35 & 0.53 & 10.11 & 0.00 & 10.11 &  9.58 & 1213 & 38 & 0.031 \\
 \\
SV Hya &  -1.70 & 0.43 & 10.49 & 0.34 & 10.15 &  9.72 & 1138 & 43 & 0.038 \\
WZ Hya &  -1.30 & 0.55 & 10.85 & 0.26 & 10.59 & 10.04 &  982 & 51 & 0.052 \\
V Ind  & -1.50 & 0.49 &  9.93 & 0.05 &  9.88 &  9.39 & 1324 & 33 & 0.025 \\
RR Leo &  -1.57 & 0.47 & 10.72 & 0.09 & 10.63 & 10.16 &  929 & 48 & 0.052 \\
U Lep &   -1.93 & 0.36 & 10.58 & 0.02 & 10.56 & 10.20 &  912 & 46 & 0.050 \\
 \\
RR Lyr &  -1.37 & 0.53 &  7.66 & 0.13 &  7.53 &  7.00 & 3981 & 30 & 0.008 \\
CN Lyr &  -0.26 & 0.86 & 11.46 & 0.62 & 10.84 &  9.98 & 1009 & 66 & 0.065 \\
KX Lyr &  -0.46 & 0.80 & 11.00 & 0.14 & 10.86 & 10.06 &  973 & 56 & 0.057 \\
RV Oct &  -1.34 & 0.54 & 10.95 & 0.35 & 10.60 & 10.06 &  973 & 52 & 0.053    \\
UV Oct &  -1.61 & 0.46 &  9.43 & 0.21 &  9.22 &  8.76 & 1770 & 39 & 0.022  \\
 \\
AV Peg &  -0.14 & 0.90 & 10.48 & 0.14 & 10.34 &  9.44 & 1294 & 43 & 0.033   \\
BH Peg &  -1.38 & 0.53 & 10.44 & 0.20 & 10.24 &  9.71 & 1143 & 42 & 0.037 \\
AR Peg &  -0.43 & 0.81 & 10.45 & 1.08 &  9.37 &  8.56 & 1941 & 42 & 0.022 \\
HH Pup &  -0.69 & 0.73 & 11.23 & 0.35 & 10.88 & 10.15 &  933 & 61 & 0.065  \\
V440 Sgr & -1.47 & 0.50 & 10.30 & 0.36 &  9.94 &  9.44 & 1294 & 40 & 0.031 \\
 \\
RU Scl &  -1.25 & 0.56 & 10.17 & 0.03 & 10.14 &  9.58 & 1213 & 39 & 0.032  \\
VY Ser &  -1.82 & 0.39 & 10.14 & 0.06 & 10.08 &  9.69 & 1153 & 38 & 0.033 \\
AN Ser &  -0.04 & 0.93 & 11.00 & 0.09 & 10.91 &  9.98 & 1009 & 55 & 0.055  \\
RW TrA &  +0.07 & 0.96 & 11.33 & 0.31 & 11.02 & 10.06 &  973 & 65 & 0.067 \\
RV UMa &  -1.19 & 0.58 & 10.66 & 0.01 & 10.65 & 10.07 &  968 & 48 & 0.050 \\
 \\
TU UMa &  -1.44 & 0.51 &  9.83 & 0.00 &  9.83 &  9.32 & 1368 & 33 & 0.024  \\
UU Vir &  -0.82 & 0.69 & 10.57 & 0.01 & 10.56 &  9.87 & 1062 & 47 & 0.044  \\
\tableline 
 &  \\ 
\multicolumn{10}{c}{1.01 kpc $<$ D(Layden) $<$ 1.25 kpc} \\ 
 & \\ 
TY Aps &  -1.21 & 0.58 & 11.75 & 0.55 & 11.20 & 10.62 &  752 & 78 & 0.104 \\
SW Aqr &  -1.24 & 0.57 & 11.14 & 0.22 & 10.92 & 10.35 &  851 & 60 & 0.071   \\
DN Aqr &  -1.63 & 0.45 & 11.18 & 0.02 & 11.16 & 10.71 &  721 & 60 & 0.083 \\
MS Ara &  -1.48 & 0.50 & 11.29 & 0.30 & 10.99 & 10.49 &  798 & 62 & 0.078 \\
ST Boo &  -1.86 & 0.38 & 11.01 & 0.04 & 10.97 & 10.59 &  762 & 57 & 0.075 \\
 \\
TW Boo &  -1.41 & 0.52 & 11.25 & 0.01 & 11.24 & 10.72 &  718 & 60 & 0.083 \\
UY Boo &  -2.49 & 0.19 & 10.91 & 0.00 & 10.91 & 10.72 &  718 & 51 & 0.071 \\
TT Cnc &  -1.58 & 0.47 & 11.33 & 0.12 & 11.21 & 10.74 &  711 & 65 & 0.091 \\
RV Cap &  -1.72 & 0.42 & 10.99 & 0.11 & 10.88 & 10.46 &  809 & 55 & 0.068 \\
V499 Cen & -1.56 & 0.47 & 11.05 & 0.18 & 10.87 & 10.40 &  832 & 58 & 0.070 \\
 \\
RY Col &  -1.11 & 0.61 & 10.90 & 0.01 & 10.89 & 10.28 &  879 & 52 & 0.059 \\
BK Dra &  -2.12 & 0.30 & 11.34 & 0.18 & 11.16 & 10.86 &  673 & 60 & 0.089 \\
SX For &  -1.62 & 0.45 & 11.08 & 0.00 & 11.08 & 10.63 &  748 & 57 & 0.076 \\
RR Gem &  -0.35 & 0.84 & 11.42 & 0.21 & 11.21 & 10.37 &  843 & 60 & 0.071 \\
TW Her &  -0.67 & 0.74 & 11.29 & 0.17 & 11.12 & 10.38 &  839 & 58 & 0.069 \\
 \\
 \\
SZ Hya &  -1.75 & 0.42 & 11.25 & 0.05 & 11.20 & 10.78 &  698 & 60 & 0.086 \\
SS Leo &  -1.83 & 0.39 & 11.07 & 0.04 & 11.03 & 10.64 &  745 & 56 & 0.075 \\
RZ Leo &  -2.13 & 0.30 & 11.43 & 0.32 & 11.11 & 10.81 &  689 & 58 & 0.084  \\
RY Oct &  -1.83 & 0.39 & 11.35 & 0.31 & 11.04 & 10.65 &  741 & 57 & 0.077 \\
CG Peg &  -0.48 & 0.80 & 11.18 & 0.20 & 10.98 & 10.18 &  920 & 56 & 0.061 \\
 \\
U Pic  & -0.73 & 0.72 & 11.38 & 0.00 & 11.38 & 10.66 &  738 & 65 & 0.088 \\
AV Ser &  -1.20 & 0.58 & 11.52 & 0.26 & 11.26 & 10.68 &  731 & 60 & 0.082 \\
AB UMa &  -0.72 & 0.72 & 11.14 & 0.00 & 11.14 & 10.42 &  824 & 59 & 0.072 \\
\tableline 
 &  \\
\multicolumn{10}{c}{1.26 kpc $<$ D(Layden) $<$ 1.40 kpc} \\
 & \\ 
BR Aqr &  -0.84 & 0.69 & 11.45 & 0.04 & 11.41 & 10.72 &  718 & 68 & 0.095 \\
AA Aql &  -0.58 & 0.77 & 11.77 & 0.21 & 11.56 & 10.79 &  695 & 80 & 0.115 \\
V674 Cen & -1.53 & 0.48 & 11.65 & 0.18 & 11.47 & 10.99 &  634 & 76 & 0.120 \\
RX Cet &  -1.46 & 0.50 & 11.43 & 0.03 & 11.40 & 10.90 &  661 & 68 & 0.103 \\
W Crt &   -0.50 & 0.79 & 11.53 & 0.09 & 11.44 & 10.65 &  741 & 72 & 0.097 \\
 \\
VW Dor &  -1.24 & 0.57 & 11.72 & 0.18 & 11.54 & 10.97 &  640 & 78 & 0.122 \\
BC Dra &  -2.00 & 0.34 & 11.57 & 0.18 & 11.39 & 11.05 &  617 & 72 & 0.117 \\
BB Eri &  -1.51 & 0.49 & 11.52 & 0.03 & 11.49 & 11.00 &  631 & 71 & 0.113 \\
VZ Her &  -1.03 & 0.63 & 11.49 & 0.12 & 11.37 & 10.74 &  711 & 68 & 0.096 \\
ST Leo &  -1.29 & 0.55 & 11.48 & 0.09 & 11.38 & 10.83 &  682 & 68 & 0.100 \\
 \\
Z Mic &   -1.28 & 0.56 & 11.60 & 0.23 & 11.37 & 10.81 &  689 & 73 & 0.106 \\
SS Oct &  -1.60 & 0.46 & 11.61 & 0.23 & 11.38 & 10.92 &  655 & 73 & 0.112 \\
V413 Oph & -1.00 & 0.64 & 11.74 & 0.63 & 11.11 & 10.47 &  805 & 79 & 0.098 \\
W Tuc &   -1.64 & 0.45 & 11.41 & 0.00 & 11.41 & 10.96 &  643 & 67 & 0.104 \\
ST Vir &  -0.88 & 0.68 & 11.57 & 0.07 & 11.50 & 10.82 &  685 & 71 & 0.104 \\
 \\
AF Vir &  -1.46 & 0.50 & 11.52 & 0.01 & 11.51 & 11.01 &  628 & 71 & 0.113 \\
AM Vir &  -1.45 & 0.50 & 11.48 & 0.14 & 11.34 & 10.84 &  679 & 68 & 0.100 \\
AT Vir &  -1.91 & 0.37 & 11.33 & 0.04 & 11.29 & 10.92 &  655 & 65 & 0.099 \\
\enddata 
\end{deluxetable}
\newpage
\clearpage
%
%
\begin{deluxetable}{lcccc}
\tablecaption{ Maximum Likelihood Simulated Absolute Magnitudes for the Layden Sample}
\tablewidth{0pt}
\tablehead{
 Subsample &  N  &  $\langle M \rangle$  & $ <M> + 1 \sigma$ &  $<M> - 1 \sigma$ \\
   (1)       &           (2)    &    (3)    &   (4)     &     (5) \\
}
\startdata
\multicolumn{5}{c}{\bf Assumed true absolute magnitude = +0.60} \\
 & \\
 all stars                        &  82   &   0.608  &  0.613   &   0.603 \\
$ {\rm [Fe/H]} \geq -1.0 $        &  23   &   0.595  &  0.612   &   0.578 \\
$ -1.0 > {\rm [Fe/H]} \geq -1.5 $   &  29   &   0.610  &  0.615   &   0.605 \\
$ -1.5 > {\rm [Fe/H]} $           &  31   &   0.613  &  0.624   &   0.602 \\
 & \\  
$ V_{0} < 10.00 $                 &  16   &   0.601  &  0.605   &   0.595 \\
$ 10.00 \leq V_{0} < 11.00 $      &  33   &   0.581  &  0.598   &   0.563 \\
$ 11.00 \leq V_{0} < 12.00 $      &  33   &   0.677  &  0.710   &   0.644 \\
\enddata
\end{deluxetable}
%
\clearpage
%
%
\begin{deluxetable}{ccc}
\tablecaption{ Bias and Scatter due to Intrinsic Spread in Absolute 
Magnitudes for Maximum Likelihood Method applied to the Layden Sample}
\tablewidth{0pt}
\tablehead{
 Intrinsic $\sigma (M)$ & Inferred $\langle M \rangle$  &  Rms $\langle M \rangle$  \\ 
   (1)       &           (2)    &    (3)  \\
}
\startdata
\multicolumn{3}{c}{\bf Assumed true absolute magnitude = +0.60 } \\
 \\
\multicolumn{3}{c}{ a) Full Layden sample of 82 objects} \\
0.00   &  0.600  &  0.006 \\ 
0.05   &  0.601  &  0.022 \\ 
0.10   &  0.604  &  0.044 \\
0.15   &  0.607  &  0.065 \\
0.20   &  0.611  &  0.087 \\
0.30   &  0.622  &  0.130 \\
0.40   &  0.637  &  0.174 \\
0.50   &  0.656  &  0.218 \\
 \\
\multicolumn{3}{c}{ b) 23 objects with ${\rm [Fe/H]} \geq -1.0$ } \\
0.00   &  0.601  &  0.018 \\
0.05   &  0.605  &  0.026 \\
0.10   &  0.610  &  0.039 \\
0.15   &  0.616  &  0.055 \\
0.20   &  0.623  &  0.071 \\
0.30   &  0.640  &  0.104 \\
0.40   &  0.661  &  0.138 \\
0.50   &  0.686  &  0.173 \\
\enddata
\end{deluxetable}
%
%
\end{document}